\documentclass[runningheads]{llncs}
\usepackage[T1]{fontenc}

\usepackage{stmaryrd}
\usepackage{amssymb}
\usepackage{amsmath}
\usepackage{braket}

\usepackage{xcolor}
\usepackage{graphicx}
\usepackage{subcaption}
\usepackage[hidelinks]{hyperref}

\usepackage{cleveref}

\newcommand{\twrchanged}[1]{#1}

\usepackage{graphicx}
\usepackage{subcaption}
\usepackage{booktabs}
\usepackage{siunitx}
\usepackage{multirow}
\usepackage{tabularx}

\newcommand{\TO}{\textsc{to}} % timeout
\makeatletter
\renewcommand\paragraph{%
  \@startsection{paragraph}{4}{\parindent}%
    {0pt}{-0.6em}%
    {\normalfont\normalsize\itshape}%
}%
\makeatother

\begin{document}
\title{\textsc{Manjushri}: A Tool for Equivalence Checking \\
of Quantum Circuits}

\author{
Xuan Du Trinh\inst{1}\orcidID{0009-0009-5610-462X} \and
Meghana Sistla\inst{2}\orcidID{0000-0002-4215-0651} \and
Nengkun Yu\inst{1}\orcidID{0000-0003-1188-3032} \and
Thomas Reps\inst{3}\orcidID{0000-0002-5676-9949}
}

\institute{
Stony Brook University, Stony Brook, NY, USA 
\and
University of Texas at Austin, Austin, TX, USA
\and
University of Wisconsin–Madison, Madison, WI, USA
}

\authorrunning{X. Trinh et al.}

%

\maketitle              % typeset the header of the contribution
\begin{abstract}
Verifying whether two quantum circuits are equivalent is a central challenge in the compilation and optimization of quantum programs. We introduce \textsc{Manjushri},\footnote{Manjushri, the Bodhisattva of Wisdom and the Blade, symbolizes sharp, analytical, and transcendent insight that severs false perceptions, attachments, and the mental barriers that cause emotional instability, while discerning even the subtlest differences.} a new automated framework for scalable quantum-circuit equivalence checking. \textsc{Manjushri} uses local projections as discriminative circuit fingerprints, implemented
with weighted binary decision diagrams (WBDDs), yielding a compact and efficient symbolic representation of quantum behavior.

\hspace{1.5ex}
We present an extensive experimental evaluation that, for random 1D Clifford+$T$ circuits, explores the trade-off between \textsc{Manjushri} and
\textsc{ECMC}, a tool for equivalence checking based on a much different approach.
\textsc{Manjushri} is much faster up to depth 30 (with the crossover point varying from 39--49, depending on the number of qubits and whether the input circuits are equivalent or inequivalent):
when inputs are equivalent, \textsc{Manjushri} is about 10$\times$ faster (or more);
when inputs are inequivalent, \textsc{Manjushri} is about 8$\times$ faster (or more).
For both kinds of equivalence-checking outcomes, \textsc{ECMC}'s success rate out to depth 50 is impressive on 32- and 64-qubit circuits:
on such circuits, \textsc{ECMC} is almost uniformly successful.
However, \textsc{ECMC} struggled on 128-qubit circuits for some depths.
\textsc{Manjushri} is almost uniformly successful out to about depth 38, before tailing off to about 75\% at depth 50 (falling to 0\% at depth 48 for 128-qubit circuits that are equivalent).

\hspace{1.5ex}
These results establish that \textsc{Manjushri} is a practical and scalable solution for large-scale quantum-circuit verification, and would be the preferred choice unless clients need to check equivalence of circuits of depth $>$38.

\end{abstract}
\section{Introduction}
\label{Se:Introduction}

%\twr{We should be consistent about terminology.  In the OOPSLA paper, we used the term ``Projection-Based Equivalence Checking (PBEC),'' so we should use that term when talking about the technique, and \textsc{Manjushri} when talking about the tool.At some point we had a variety of terms for the technique.}\ny{Good point.}

% \twr{
% Another issue about consistent terminology: instead of talking about “equivalence checking” and “inequivalence checking,” we need to change things throughout (in the text, and in the headings of the plots) to say “equivalent inputs” and “inequivalent inputs”.
% (It is misleading to say “equivalence checking” and “inequivalence checking”;
% it makes it sound like there are two algorithms.
% }

Program equivalence is a central concept in computer science, with broad applications in software engineering, compiler validation, optimization, and program analysis \cite{BERGSTRA1982113,Benton2004,10.1145/358438.349314}. Techniques for reasoning about program behavior have been extended from deterministic programs to probabilistic and approximate settings, addressing uncertainty and noise in practical systems \cite{Barthe2009,Barthe13}.

With the rise of quantum programming languages, the formal verification of quantum programs has received increasing attention \cite{ying2010predicate,Ying16,rand2018qwire,Peng2022}. In particular, a growing body of work focuses on reasoning about the equivalence of quantum programs and circuits \cite{Barthe19,unruh2019,unruh2021,Yan2024Approximate,10.1145/3763153}.

Beyond logic-based approaches, computational methods for quantum-circuit equivalence checking have attracted substantial attention. Direct classical simulation of quantum circuits is generally infeasible because the quantum state space grows exponentially with the number of qubits. Formally, equivalence checking requires proving that the quantum channels induced by two circuits are identical for all possible input states. This task is particularly challenging because, although the underlying Hilbert space is finite-dimensional, its dimension is exponential in the number of qubits and the set of input states has infinite cardinality. Together, these factors make naive state-space enumeration or exhaustive testing fundamentally intractable.
%\twr{I would say that the state space has finite dimensionality (which grows exponentially in the number of qubits), but infinite \emph{cardinality}.} \ny{Yes.}
Existing methods provide strong theoretical guarantees for exact verification, but they neither scale to large circuits nor support efficient approximate reasoning. These limitations motivate the development of automated and scalable verification tools.

Recent work has begun exploring scalable methods for quantum circuit equivalence checking, including approaches based on weighted model counting (WMC)~\cite{Mei2024Equivalence} and
collections of local projections as constraints
\cite{10.1145/3763153}.
The WMC-based approach encodes the behavior of quantum circuits into weighted Boolean formulas, reducing equivalence checking to a sequence of weighted model-counting problems. This method, implemented in tools such as \textsc{ECMC} \cite{Mei2024Equivalence}, supports universal gate sets, including non-Clifford gates like Toffoli, and can exploit parallelism for large-scale verification.

Projection-Based Equivalence Checking (PBEC) \cite{10.1145/3763153},
on the other hand, aims to reduce verification complexity by exploiting the structure of shallow circuits. A major difficulty in understanding and verifying quantum circuits arises from multipartite entanglement, which correlates many qubits in ways that are difficult to analyze globally. The intuition behind PBEC is that the depth of a quantum circuit not only represents its running time but also influences the rate at which multipartite entanglement is generated. In shallow circuits, entanglement spreads only partially, so the global output state can often be well-approximated by its marginal distributions over small subsets of qubits. This observation suggests that equivalence checking of shallow circuits may be achievable by reasoning over these low-order marginals rather than the full infinite state space of exponential dimension.

Despite these promising directions, current methods still struggle when scaling to circuits with hundreds of qubits or significant depth. WMC-based methods face combinatorial explosion, and 
\twrchanged{
PBEC
}
become less effective as entanglement spreads across larger subsystems. As a result, efficiently handling large, deep circuits remains an open challenge, highlighting a critical gap in practical
\twrchanged{
techniques for verifying quantum-circuit equivalence
}
for near-term and future quantum computing devices.

In this paper, we present \textsc{Manjushri}, an automated framework for scalable
\twrchanged{
equivalence checking of quantum circuits via PBEC.
\textsc{Manjushri} uses weighted binary decision diagrams (WBDDs) to represent local projections, resulting in a compact and efficient symbolic representation of a quantum circuit's behavior.
}

\twrchanged{
The experiments with \textsc{Manjushri} reported in \Cref{Se:Experiments} complement the experiments in the original paper on PBEC \cite{10.1145/3763153}.
In that paper, the benchmarks are 1D random circuits, where every
2-qubit gate is generated from a Haar distribution (created for a specified number of qubits $n$ and a specified depth $d$).
For checking equivalence,\footnote{
  What we call ``equivalence'' in this paper was called ``strong equivalence'' in \cite{10.1145/3763153}.
}
circuit sizes were quite limited:
$12 \leq n \leq 100$ with depth $d = 3$.
In the present paper, the benchmarks are random 1D Clifford+$T$ circuits\footnote{
\twrchanged{
  Our experiments check pairs of equivalent circuits (checking a given circuit $C$ and $\textrm{Optimize}(C)$) and pairs of inequivalent circuits (checking C and $\textrm{Mutate}(\textrm{Optimize}(C))$).
  In both kinds of experiments, \textsc{Manjushri} must handle more than Clifford+$T$ circuits because even if $C$ is a Clifford+$T$ circuit, with the optimizer we are using, PyXZ \cite{QPL:kissinger2019pyzx}, the circuit produced by $\textrm{Optimize}(C)$ can have $R_z$ and $CZ$ gates in it.
}
}
and much larger: $n \in \{ 32, 64, 128 \}$ and $1 \leq d \leq 50$.
Moreover, \Cref{Se:Experiments} gives a detailed comparison of the relative performance of \textsc{Manjushri} and the \textsc{ECMC} tool (which performs equivalence checking via model counting).
}

\section{Background and Related Work}

\paragraph{Qubits.}
A \emph{qubit} is the basic unit of quantum information. In the computational basis $\{\ket{0},\ket{1}\}$, a qubit state is $\ket{\psi}=\alpha\ket{0}+\beta\ket{1}$, where $\alpha,\beta\in\mathbb{C}$ and $|\alpha|^2+|\beta|^2=1$. Measurement yields outcomes $\ket{0}$ and $\ket{1}$ with probabilities $|\alpha|^2$ and $|\beta|^2$, respectively.

\paragraph{Quantum States.}
A pure quantum state is a unit vector $\ket{\psi}$ in a Hilbert space that encodes the probabilistic behavior of a system. Quantum states may be expressed as superpositions of basis states.

\twrchanged{
\paragraph{Tensor Product and Multi-Qubit Systems.}
For an $n$-qubit system, the joint state space is constructed as the tensor (Kronecker) product of $n$ two-dimensional qubit spaces, yielding a Hilbert space of dimension $2^n$---i.e., the dimension grows exponentially with the number of qubits $n$.
This exponential growth in dimensionality enables the representation of complex multi-qubit states, including entangled states, and forms the mathematical foundation for modeling multi-qubit quantum gates.
}

\paragraph{Unitary Operations.}
Quantum evolution is described by unitary operators $U$ satisfying $U^\dagger U = U U^\dagger = I$.
This condition ensures reversibility and preserves state norms. Unitary operators acting on one or two qubits are called \emph{quantum gates}.

\twrchanged{
The tensor (Kronecker) product combines operators that act on individual subsystems into a single operator on the composite system.
}

\paragraph{Quantum Circuits.}
Quantum circuits define unitary transformations over an $n$-qubit Hilbert space through a sequence of gate layers. Each layer consists of gates acting on disjoint subsets of qubits and can therefore be executed in parallel. For clarity, we focus on circuits composed of one- and two-qubit gates; all results extend directly to gates acting on at most a fixed constant number $m$ of qubits.

Let $U_i^{(k)}$ denote the gate applied to qubit(s) indexed by $i$ in layer $k$. The unitary operator corresponding to layer $k$ is defined as $U^{(k)} = \bigotimes_i U_i^{(k)}$.
A circuit of depth $L$ is interpreted as the sequential composition of these layer operators, yielding the overall transformation
\begin{equation}
\label{Eq:QuantumCircuitSemantics}
\ket{\psi_{\mathrm{out}}} = U^{(L)} U^{(L-1)} \cdots U^{(1)} \ket{\psi_{\mathrm{in}}}.
\end{equation}
This formulation captures the operational semantics of a quantum circuit as a product of layer-wise unitary transformations. A unitary $U$ induces a quantum operation on density matrices by conjugation. For a pure state $\rho = \ket{\psi}\!\bra{\psi}$, we have $\mathcal{U}(\rho) := U \rho U^\dagger = U \ket{\psi}\!\bra{\psi} U^\dagger$,
where we define $\mathcal{U} := \lambda x.\, U x U^\dagger$. This coincides with the standard evolution of pure states.

\paragraph{Circuit Equivalence.}
Quantum states that differ by a global phase factor $e^{i\theta}$ are physically indistinguishable. 
This property motivates the notion of equivalence.

\begin{definition}[Circuit equivalence]
Two circuits $C_0 = \prod_{k=1}^{L} \bigotimes_{i} U_{i}^{(k)},$ and $C_1 = \prod_{k=1}^{T} \bigotimes_{j} V_{j}^{(k)}$ are \emph{equivalent} if, for every input state $n$-qubit $\ket{\psi}$, their output states are equal up to a global phase.
\end{definition}

%\paragraph{Related Work: \textsc{ECMC}.}
%\cite{Mei2024Equivalence} reduces quantum circuit equivalence checking to weighted model counting (WMC) via the \textsc{ECMC} tool. Circuit behavior is encoded in the Pauli basis, and equivalence is checked through $2n$ parallel WMC queries using GPMC, supporting universal gate sets including non-Clifford gates like Toffoli. Evaluated against QCEC on random Clifford+T circuits with a 300-second timeout, \textsc{ECMC} outperforms QCEC on non-equivalent or heavily non-Clifford circuits, while QCEC excels on structured or equivalent cases, with runtimes up to 300~s for 90-qubit circuits of depth 20.

\subsection*{Related Work}

\paragraph{Scalable Equivalence Checking via Local Projections.}
\twrchanged{
Yu et al.\ \cite{10.1145/3763153} focused on shallow (constant-depth) quantum circuits and introduced
}
Projection-Based Equivalence Checking (PBEC), which constructs a set of local-projection constraints that uniquely characterize a circuit’s output state, enabling equivalence checking without full state simulation.
\twrchanged{
For circuits of fixed depth, the PBEC algorithm for checking equivalence runs
}
in time and space linear in the number of qubits, exploiting the fact that a small set of local constraints suffices to capture the output state. Beyond equivalence checking, this representation also supports sound and complete assertion checking for properties expressed as conjunctions of local projections. Experimentally, equivalence checking between 100-qubit circuits of depth 3 took a few seconds to tens of seconds, and computing the constraint description for a 100-qubit circuit of depth 6 required under 20 seconds on large-memory multi-core hardware, demonstrating the practical scalability of the approach.

\paragraph{Equivalence Checking of Quantum Circuits via Model Counting.}
\twrchanged{Mei et al.}~\cite{Mei2024Equivalence} reduces the $n$-qubit circuit equivalence problem, $\lambda x.\, U x U^\dagger = \lambda x.\, V x V^\dagger$, to $2n$ matrix-equality conditions, $\lambda x.\, U x U^\dagger\, P_i = \lambda x.\, V x V^\dagger \,  P_i$, where each $P_i$ is a single-qubit Pauli matrix $X$ or $Z$ and each condition involves two $2^n \times 2^n$ matrices.
The \textsc{ECMC} tool extends Pauli coefficient updates from Clifford circuits to Clifford+T+Toffoli circuits when appending a new gate layer and encodes these coefficients as weighted model counts, summing the weights of all satisfying assignments of a Boolean formula.

\twrchanged{
\paragraph{Quasimodo.}
Quasimodo\footnote{
   \url{https://github.com/trishullab/Quasimodo}
} \cite{sistla2023symbolic} is an open-source Python library for symbolic simulation of quantum circuits, which supports different kinds of decision diagrams as the underlying data structure for representing gates, unitary matrices, etc.---in particular, CFLOBDDs~\cite{sistla2024cflobdds}, WCFLOBDDs~\cite{sistla2024weighted}, BDDs (specifically CUDD \cite{ICCAD:BFGHMPS93}), and Weighted BDDs (WBDDs, specifically MQTDD~\cite{zulehner2019efficiently}).
\textsc{Manjushri} uses Quasimodo to implement the PBEC algorithm \cite{10.1145/3763153}.
For the experiments described in \Cref{Se:Experiments}, \textsc{Manjushri} was instantiated with WBDDs as the back-end representation;
we found that for PBEC, the other kinds of decision diagrams supported by Quasimodo were not competitive.
}

\section{Experiments}
\label{Se:Experiments}

\subsection{Experimental Setup}

\paragraph{(1) Compared tools and methods.}
We benchmarked \textsc{ECMC} \cite{Mei2024Equivalence} against \textsc{Manjushri},\footnote{
\twrchanged{
  \textsc{Manjushri} is available as the ``equivalence'' branch of Quasimodo (\url{https://github.com/trishullab/Quasimodo/tree/equivalence}).
  %%   \textsc{Manjushri} is available at \twr{MISSING: give URL}. 
}
} our implementation of
\twrchanged{
PBEC
}
\cite{10.1145/3763153}.
\twrchanged{
\textsc{Manjushri} uses the \textsc{MQTDD}\footnote{
\twrchanged{
  We made minor modifications to MQTDD to accommodate the operations used in PBEC.
}
}
data structure \cite{zulehner2019efficiently} (essentially an implementation of WBDDs), for representing and manipulating circuit semantics.
}
All input circuits are provided in QASM format. While \textsc{ECMC} operates directly on the QASM inputs, \textsc{Manjushri} first converts each QASM circuit into an \textsc{MQTDD}-based internal representation. Accordingly, unless stated otherwise, the reported runtime for \textsc{Manjushri} is defined as
$$T_{\textsc{Manjushri}} = T_{\text{convert}} + T_{\text{check}},$$
where $T_{\text{convert}}$ denotes QASM-to-\textsc{MQTDD} conversion time and $T_{\text{check}}$ denotes equivalence-checking time on the converted representation. The reported runtime for \textsc{ECMC} corresponds to its equivalence-checking procedure on the original QASM inputs. All runs enforce a per-instance timeout of 100 seconds. The equivalence tolerance is set to $1\times 10^{-15}$ for both tools: \textsc{Manjushri} uses this value as its numerical comparison tolerance between density matrices (see \cite{10.1145/3763153}), and \textsc{ECMC} uses it as the floating-point-error (FPE) threshold in the backend solver configuration.

\twrchanged{
The experiments were run on a server equipped with 2× Intel\textsuperscript{\textregistered} Xeon\textsuperscript{\textregistered} Gold 6338 CPUs (128 threads total) and 1.0~TiB of RAM, running Ubuntu 20.04.6 LTS.
}

\paragraph{(2) Random 1D Clifford+$T$ circuits (equivalence: original vs.\ optimized).}
We generate random
\twrchanged{
1D
}
Clifford+$T$ circuits with the following gate-density profile (percent of gate placements): $H$ 35\%, $S$ 35\%, $\mathrm{CNOT}$ 10\%, and $T$ 20\% (no identity gates),
\twrchanged{
the same percentages used by Mei et al.\ \cite{Mei2024Equivalence}.
}
Two-qubit gates are nearest-neighbor $\mathrm{CNOT}$s on adjacent qubits, while single-qubit gates are sampled from $\{H,S,T\}$ according to the specified proportions. We evaluate circuits with $n \in \{32,64,128\}$ qubits.

For each configuration $(n,\mathrm{depth})$, we generate $100$ random circuits (``original'') and apply \textsc{PyZX} optimization to obtain corresponding ``optimized'' circuits, resulting in $100$ original/optimized pairs per $(n,\mathrm{depth})$. For the $n=128$ setting, \textsc{PyZX} optimization becomes prohibitively slow at larger depths;
\twrchanged{
consequently, in this case
}
we directly optimize only circuits up to depth $7$. To construct higher-depth instances at $n=128$, we randomly sample circuit blocks of depth in $\{2,\ldots,7\}$, apply \textsc{PyZX} optimization at the block level, and then concatenate these optimized blocks until reaching the target depth, forming the corresponding high-depth original/optimized pairs. Consequently, the ``optimized'' circuits in the $n=128$ experiments are best viewed as \emph{semi-optimized} (or \emph{pseudo-optimized}) rather than globally optimized for the full target depth. This approximation does not affect fairness: both equivalence checkers are evaluated on exactly the same circuit pairs generated by the identical procedure, and the same semi-optimization pipeline is applied uniformly across all methods.

We measure the runtime of equivalence checking on each pair with a per-instance timeout of $100$ seconds. We report:
(i) \emph{completion rate}, the percentage of instances that terminate before timeout, and
(ii) \emph{success rate}, the percentage of \emph{completed} instances that return the correct equivalence outcome.

\paragraph{(3) Random Clifford+T circuits (inequivalence: optimized vs.\ mutated optimized).}
To evaluate inequivalence detection under controlled perturbations, 
\twrchanged{
we compare the original circuit against a mutated version of the optimized circuit from (2)
}
by removing one randomly chosen gate from the optimized circuit.
We then check
\twrchanged{
$\langle \textrm{original}, \textrm{mutated-optimized} \rangle$ pairs,
}
which are inequivalent by construction. All other settings match (2): the same $(n,\mathrm{depth})$ grid, $100$ pairs per configuration, a $100$-second timeout, and the same completion-rate and success-rate metrics.

% \paragraph{(4) Algorithmic circuits with injected-error variants.}
% We additionally benchmark on the algorithmic-circuit suite distributed with the \textsc{ECMC--Quokka-Sharp} \cite{Mei2024Equivalence} artifacts (the same suite used in \textsc{ECMC}'s published comparison with QCEC in  \cite{burgholzer2020advanced-QCEC}). Each original circuit is paired with multiple provided variants with known expected outcomes:
% \emph{opt} (optimized; expected equivalent),
% \emph{gm} (gate-missing; expected inequivalent),
% \emph{flip} (CNOT-flipped; expected inequivalent),
% \emph{shift4} (phase-shift of magnitude $10^{-4}$; expected inequivalent), and
% \emph{shift7} (phase-shift of magnitude $10^{-7}$; expected inequivalent).
% We evaluate both tools on these pairs under the same evaluation protocol (per-instance timeout and the same completion/success rate metrics).

\paragraph{(4) Procedure for generating random Random Clifford+$T$ circuits.}
Random Clifford+$T$ circuits are generated layer-by-layer. For each layer, qubits are scanned from left to right. At each qubit position $q$, we first attempt to place a nearest-neighbor two-qubit gate: with probability $p_{\mathrm{CNOT}}$, we apply a $\mathrm{CNOT}$ on $(q,q{+}1)$ (if $q<n{-}1$) and then skip the next qubit position (i.e., advance by two). Otherwise, with probability $p_{I}$ we place no gate on $q$ and advance by one. If neither event occurs, we place a single-qubit gate on $q$, sampled from $\{H,S,T\}$ with relative weights proportional to $(p_{H}, p_{S}, p_{T})$ (normalized so the three weights sum to $1$), and then advance by one. Repeating this procedure for the specified number of layers yields the circuit. In our experiments, we set $p_I=0$ and use the gate-density profile $p_H=0.35$, $p_S=0.35$, $p_{\mathrm{CNOT}}=0.10$, and $p_T=0.20$.

% \paragraph{(5) Input format and timing convention.}
% All benchmark circuits are provided in QASM format. \textsc{ECMC} operates directly on the QASM inputs, whereas \textsc{Manjushri} first converts each QASM circuit into its internal representation based on the \textsc{MQTDD} data structure. Since this conversion step is required for \textsc{Manjushri} across all experimental suites, we report (and time) it explicitly. Concretely, for \textsc{Manjushri} we decompose the per-instance runtime as
% \[
% T_{\textsc{Manjushri}} \;=\; T_{\text{convert}} \;+\; T_{\text{check}},
% \]
% where $T_{\text{convert}}$ denotes QASM-to-\textsc{MQTDD} conversion time and $T_{\text{check}}$ denotes the equivalence-checking time on the converted representation. For \textsc{ECMC}, the reported runtime corresponds to its equivalence-checking procedure on the original QASM inputs.

\paragraph{(5) Algorithmic circuits with injected-error variants.}
We additionally benchmark on the quantum-algorithm circuit suite distributed with the \textsc{ECMC--Quokka-Sharp} artifacts \cite{Mei2024Equivalence}, which is the same benchmark set used in \textsc{ECMC}’s published comparison with QCEC \cite{burgholzer2020advanced-QCEC}. For each original circuit, the artifact provides paired variants with known expected outcomes: \emph{opt} (optimized circuit, expected equivalent), and four inequivalence injections—\emph{gm} (gate-missing variant), \emph{flip} (CNOT control/target flipped), \emph{shift4} (phase-shift of magnitude $10^{-4}$), and \emph{shift7} (phase-shift of magnitude $10^{-7}$).

\subsection{Results and Findings}

\begin{figure}[!tb]
  \centering
  \begin{subfigure}[b]{0.49\textwidth}
    \centering
    \includegraphics[width=\linewidth]{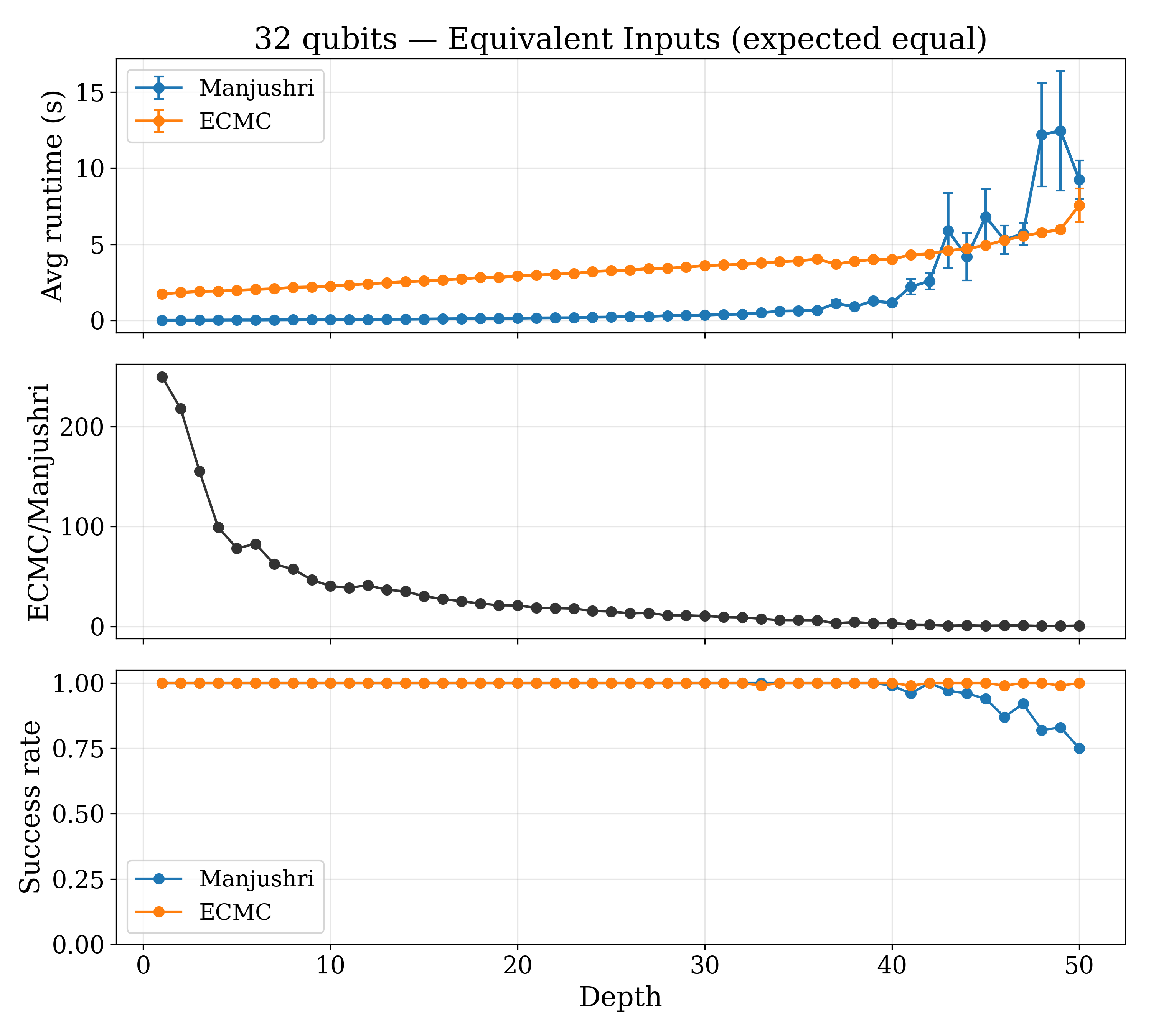}
    \caption{Equivalent inputs ($32$ qubits).}
    \label{fig:q32-equivalence}
  \end{subfigure}
  \hfill
  \begin{subfigure}[b]{0.49\textwidth}
    \centering
    \includegraphics[width=\linewidth]{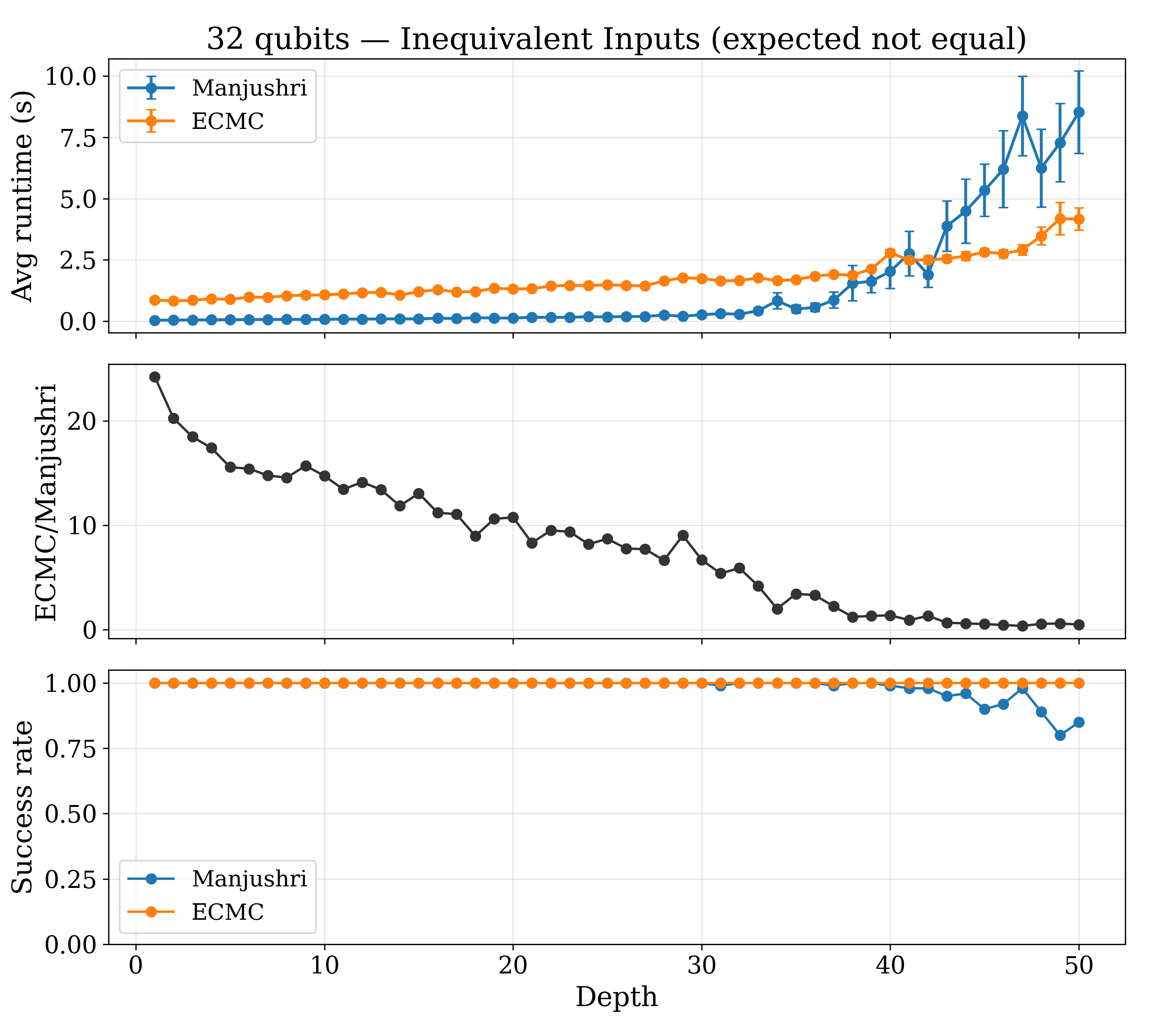}
    \caption{Inequivalent inputs ($32$ qubits).}
    \label{fig:q32-inequivalence}
  \end{subfigure}

  \caption{Results for 32-qubit random-circuit benchmarks.}
  \label{fig:q32-runtime-summary}
\end{figure}

\begin{figure}[!tb]
  \centering
  \begin{subfigure}[b]{0.49\textwidth}
    \centering
    \includegraphics[width=\linewidth]{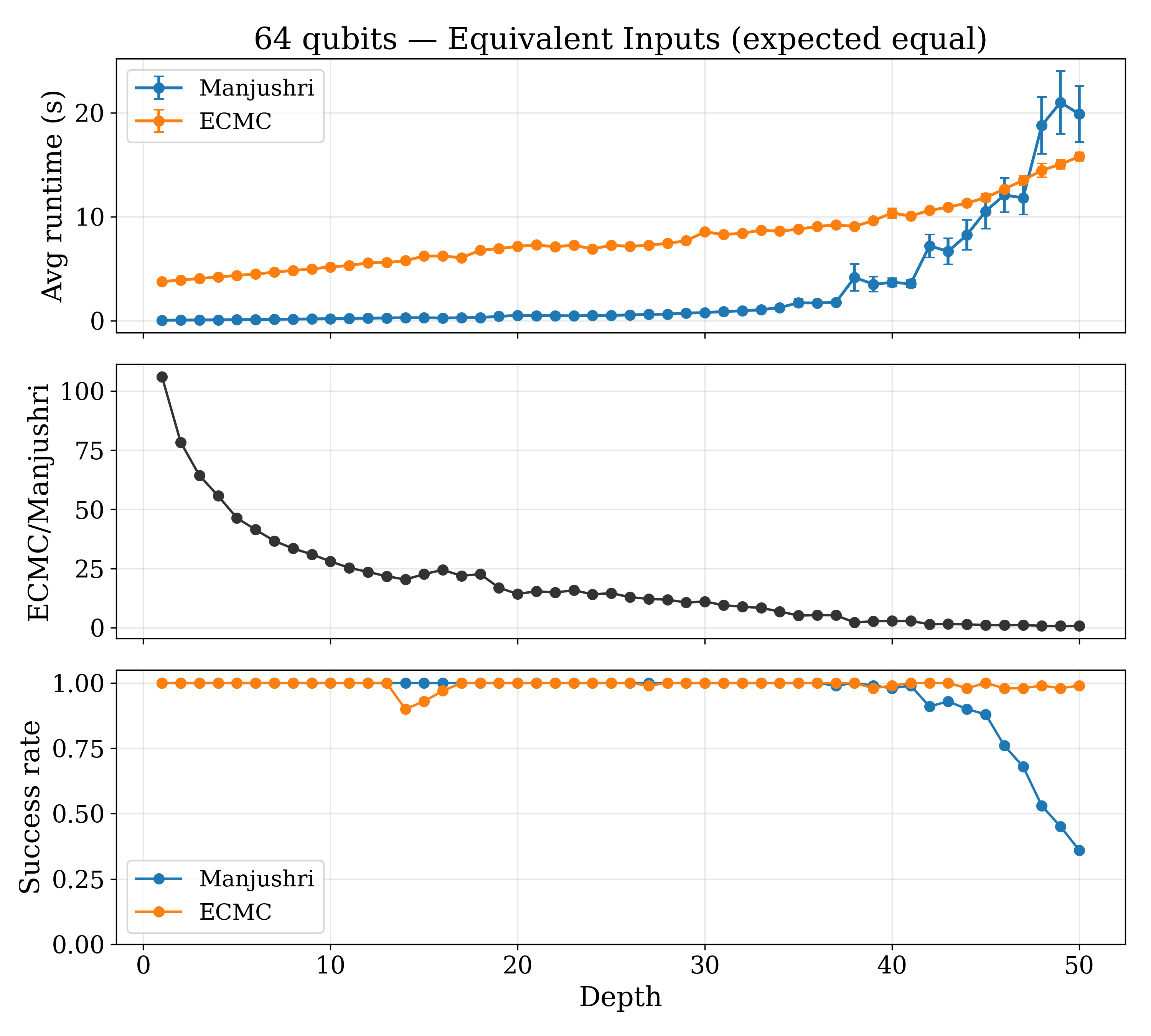}
    \caption{Equivalent inputs ($64$ qubits).}
    \label{fig:q64-equivalence}
  \end{subfigure}
  \hfill
  \begin{subfigure}[b]{0.49\textwidth}
    \centering
    \includegraphics[width=\linewidth]{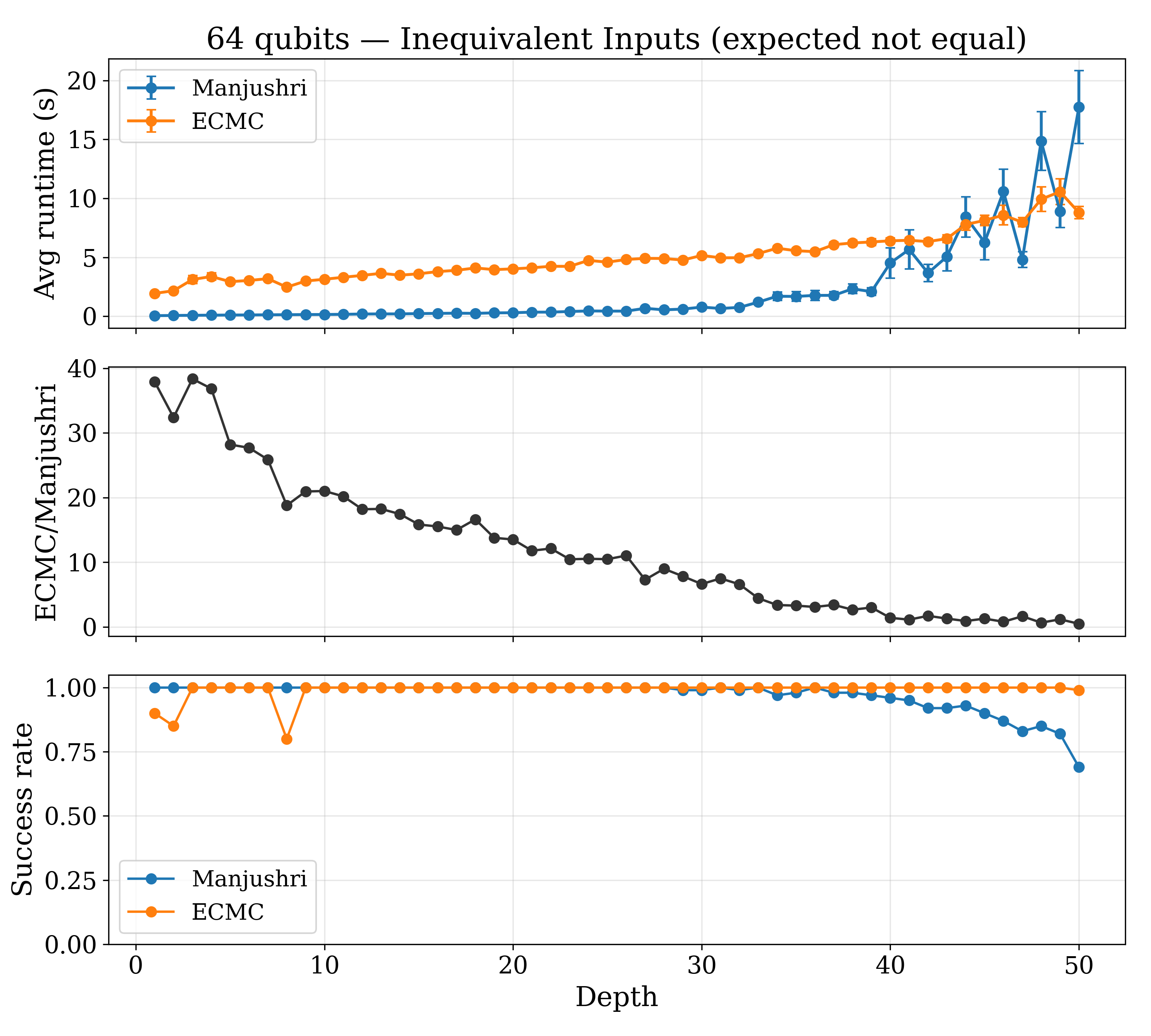}
    \caption{Inequivalent inputs ($64$ qubits).}
    \label{fig:q64-inequivalence}
  \end{subfigure}

  \caption{Results for 64-qubit random-circuit benchmarks.}
  \label{fig:q64-runtime-summary}
\end{figure}

\begin{figure}[!tb]
  \centering
  \begin{subfigure}[b]{0.49\textwidth}
    \centering
    \includegraphics[width=\linewidth]{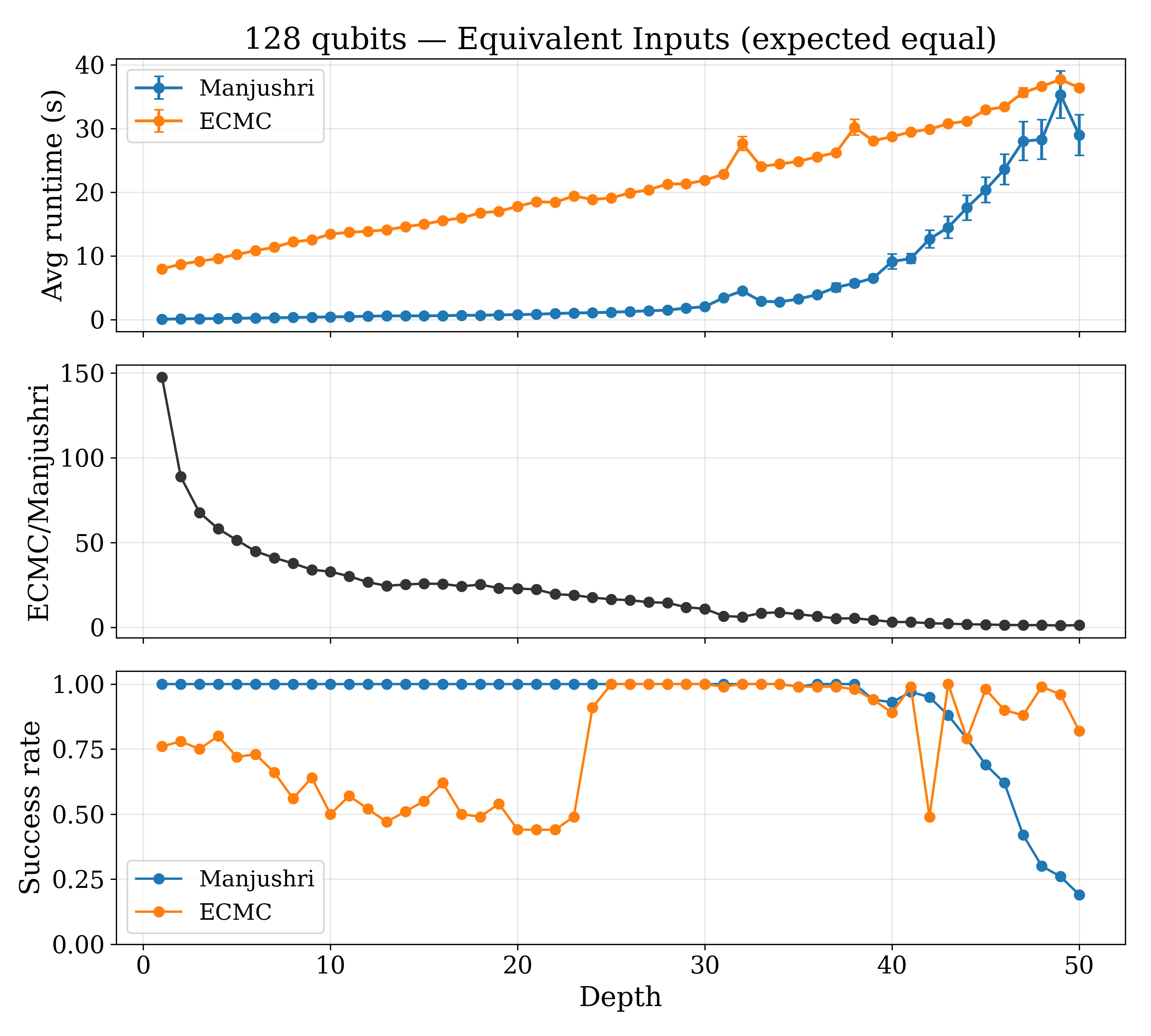}
    \caption{Equivalent inputs ($128$ qubits).}
    \label{fig:q128-equivalence}
  \end{subfigure}
  \hfill
  \begin{subfigure}[b]{0.49\textwidth}
    \centering
    \includegraphics[width=\linewidth]{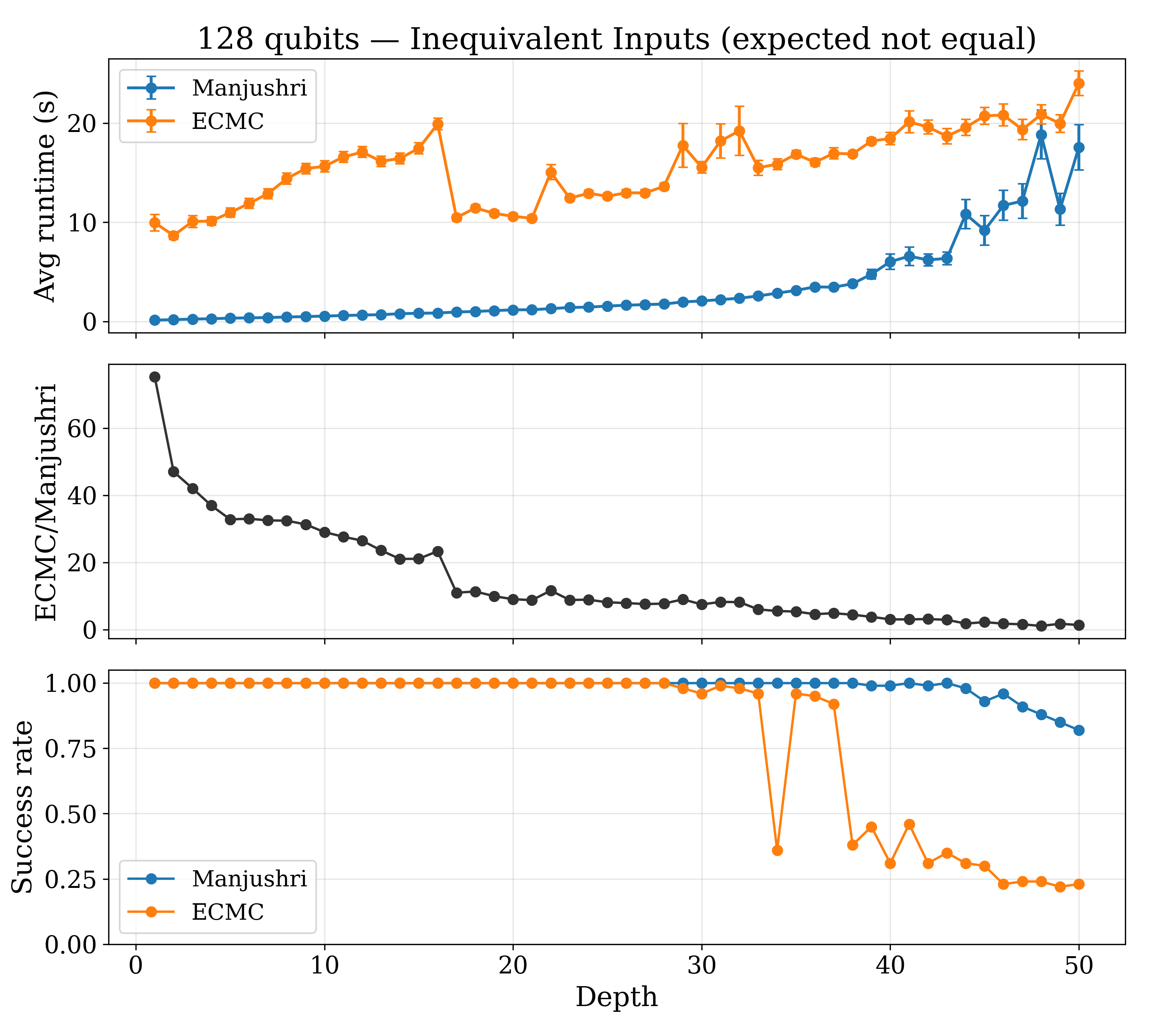}
    \caption{Inequivalent inputs ($128$ qubits).}
    \label{fig:q128-inequivalence}
  \end{subfigure}

  \caption{Results for 128-qubit random-circuit benchmarks.}
  \label{fig:q128-runtime-summary}
\end{figure}

\Cref{fig:q32-runtime-summary,fig:q64-runtime-summary,fig:q128-runtime-summary} report the results of our experiments on circuit pairs with 32, 64, and 128 qubits, respectively.
Each panel reports (top) the average runtime versus depth, with error bars showing the standard error of the mean (SEM); (middle) the runtime ratio (\textsc{ECMC} / \textsc{Manjushri}); and (bottom) the success rate.
Only runtimes for runs that do not time out and do produce correct results are counted in computing the means.

\begin{table}[!ht]
\caption{Aggregate statistics over all $5{,}000$ runs per $n_{\text{qubits}}$ and input type, obtained from the $50$ circuit depths $1 \leq d \leq 50$, with $100$ independent circuit pairs (equivalent or inequivalent) at each depth. We report the completion rate, the correctness rate (\twrchanged{percentage of correct results across runs that did not time out}), the success rate (\twrchanged{percentage of correct results across} all the runs), and the number of incorrect (non-timeout) outcomes.
%% \Du{fully updated. When I rerun the error cases of \textsc{ECMC}, the results of some flipped. But I did not changed in here.}
}
\vspace{1.0ex}
\centering
\small
\begin{tabular}{r l l r r r r}
\hline
             &       &        &            &             &         & Incorrect \\
$n_{qubits}$ & Input & Method & Completion & Correctness & Success & (non-timeout) \\
\hline\hline
32  & equivalent   & \textsc{ECMC}      & 99.98\% & 99.92\% & 99.90\% & 4 \\
64  & equivalent   & \textsc{ECMC}      & 100.00\% & 99.32\% & 99.32\% & 34 \\
128  & equivalent   & \textsc{ECMC}      & 98.02\% & 79.49\% & 77.92\% & 1005 \\
\hline
32  & inequivalent   & \textsc{ECMC}      & 100.00\% & 100.00\% & 100.00\% & 0 \\
64  & inequivalent   & \textsc{ECMC}      & 99.98\% & 99.10\% & 99.08\% & 45 \\
128  & inequivalent   & \textsc{ECMC}      & 97.82\% & 81.89\% & 80.10\% & 886 \\
\hline\hline
32  & equivalent   & \textsc{Manjushri}      & 98.32\% & 100.00\% & 98.32\% & 0 \\
64  & equivalent   & \textsc{Manjushri}      & 94.72\% & 100.00\% & 94.72\% & 0 \\
128  & equivalent   & \textsc{Manjushri}      & 91.86\% & 100.00\% & 91.86\% & 0 \\
\hline
32  & inequivalent   & \textsc{Manjushri}      & 98.36\% & 100.00\% & 98.36\% & 0 \\
64  & inequivalent   & \textsc{Manjushri}      & 96.98\% & 100.00\% & 96.98\% & 0 \\
128  & inequivalent   & \textsc{Manjushri}      & 98.60\% & 100.00\% & 98.60\% & 0 \\
\hline\hline
\end{tabular}
\label{tab:random-cliffordT-stats}
\end{table}

\Cref{tab:random-cliffordT-stats} shows aggregate statistics over all $5{,}000$ runs per $n_{\text{qubits}}$ and input type.
We found that \textsc{ECMC} does not always return correct results (see columns 5 and 7 in the rows for \textsc{ECMC}).
We contacted the authors of the \textsc{ECMC} tool for help in diagnosing the source of error(s) in \textsc{ECMC}.
They suggested changing the floating-point-error tolerance, which helps checking inequivalent inputs because it enforces harder conditions for returning the answer ``equivalent.''
Unfortunately, this change does not correct all issues: equivalence checking of \textsc{ECMC} still makes mistakes.
We will try to work with them after the submission deadline to identify and correct the root cause.

\paragraph{Findings.}
\twrchanged{
The experiments reveal that there is a trade-off between \textsc{Manjushri} and \textsc{ECMC}.
The fact that such a trade-off exists is not surprising, given that they are based on radically different techniques.
\begin{itemize}
  \item
    For circuits of 32, 64, and 128 qubits, \textsc{Manjushri} is faster for shallower circuits;
    \textsc{ECMC} is faster for deeper circuits.
    The circuit depths at the respective crossover points are as follows:

    \begin{center}
    \begin{tabular}{c|c|c}
      \multirow{3}{*}{\textbf{Qubits}} & \multicolumn{2}{c}{\textbf{Depth}}\\
                      \cline{2-3}
                      & \textbf{Equivalent} & \textbf{Inequivalent} \\
                      & \textbf{inputs}     & \textbf{inputs} \\
      \hline
      32  & 43 & 40 \\
      64  & 41 & 39 \\
      128 & 44 & 49 \\
    \end{tabular}
    \end{center}

  \item
    Regardless of whether the result is ``Equivalent'' or ``Inequivalent,'' \textsc{Manjushri} is much faster up to depth 30:
    when inputs are equivalent, \textsc{Manjushri} is about 10$\times$ faster (or more);
    when inputs are inequivalent, \textsc{Manjushri} is about 8$\times$ faster (or more).
  \item
    For both kinds of equivalence-checking outcomes, \textsc{ECMC}'s success rate out to depth 50 is impressive on 32- and 64-qubit circuits:
    on such circuits, \textsc{ECMC} is almost uniformly successful.
    However, \textsc{ECMC} struggled on 128-qubit circuits for some depths.

    \hspace{1.5ex}
    \textsc{Manjushri} is almost uniformly successful out to about depth 38, before tailing off to about 75\% at depth 50 (falling to 0\% at depth 48 for 128-qubit circuits that are equivalent).
  \item
    For both tools, answers are obtained faster when the input circuits are inequivalent than when the input circuits are equivalent.
\end{itemize}
}

\paragraph{Algorithmic circuits with injected-error variants.}
The appendix gives the results of our experiment with the benchmark suite from the \textsc{ECMC--Quokka-Sharp} artifacts\footnote{
  \url{https://github.com/System-Verification-Lab/quokka-sharp-artifacts/tree/main/benchmark/algorithm}
}
(an \textsc{ECMC} benchmark used in multiple papers from that group).

\section{Artifact Submission}
\label{Se:ArtifactSubmission}

\twrchanged{
If the paper is accepted, we intend to submit an artifact for evaluation.
}

\section{Conclusion}
\label{Se:Conclusion}

\twrchanged{
\Cref{Se:Experiments} presents an extensive experimental evaluation that, for random 1D Clifford+$T$ circuits, explores the trade-off between \textsc{Manjushri} and
\textsc{ECMC}.
}

\twrchanged{
A natural application for both \textsc{ECMC} and \textsc{Manjushri} would be for checking circuit fragments in a circuit optimizer \cite{ASPLOS:XMTA25}, or circuit-optimizer generator \cite{PLDI:XMPTA23}.
Our colleague Amanda Xu told us that both (i) the circuit fragments analyzed to create rewrite rules, and (ii) the sub-circuits that are resynthesized in her tools have no more than 3--5 qubits, and that the rewrite-rule patterns are at most depth 6 \cite{PC:AmandaXu26}.
Both \textsc{ECMC} and \textsc{Manjushri} can check equivalence of much larger circuits than that, so it might be possible to use them in an optimizer that resynthesizes larger subcircuits (with more qubits) than can currently be handled using unitary synthesis (the standard technique).
Because \textsc{Manjushri} (a) is so much faster than \textsc{ECMC} on circuits out to depth 30 (with the crossover point varying from 39--49), and (b) 
is almost uniformly successful out to about depth 38, \textsc{Manjushri} would be the preferred choice until optimizers are created that need to check equivalence of sub-circuits of depth $>$38.
}

\newpage
\bibliographystyle{splncs04}
\bibliography{main}

\newpage
\section*{Appendix}

This section give the results of our experiment with the benchmark suite from the \textsc{ECMC--Quokka-Sharp} artifacts \cite{Mei2024Equivalence}, which is the benchmark set used in \textsc{ECMC}’s published comparison with QCEC \cite{burgholzer2020advanced-QCEC}.
For each original circuit, the artifact provides paired variants with known expected outcomes: \emph{opt} (optimized circuit, expected equivalent), and four inequivalence injections—\emph{gm} (gate-missing variant), \emph{flip} (CNOT control/target flipped), \emph{shift4} (phase-shift of magnitude $10^{-4}$), and \emph{shift7} (phase-shift of magnitude $10^{-7}$).

When both \textsc{ECMC} and \textsc{Manjushri} timed out on a circuit, we do not list it in the tables in this appendix.

% Auto-assembled: all per-algorithm tables
% Generated by concatenating summary_at_least_one_complete_*.tex

% Rows where at least one method completed

\begin{table}[b]
\centering
\caption{\texttt{\detokenize{cycle10_293}}: cases where at least one method completed (incorrect results shown in bold).
}
\label{tab:cycle10-293-partial}
\scriptsize
\setlength{\tabcolsep}{3.5pt}
\renewcommand{\arraystretch}{0.95}
\begin{tabular}{p{0.40\textwidth} r r r l r r}
\toprule
Circuit & $n$ & $|G|$ & $|G^\prime|$ & variant & \textsc{ECMC} & \textsc{Manjushri} \\
\midrule
\texttt{\detokenize{cycle10_293}} & 39 & 512 & 846 & gm & 1.12 & \TO \\
\texttt{\detokenize{cycle10_293}} & 39 & 512 & 847 & opt & \TO & 21.42 \\
\bottomrule
\end{tabular}
\end{table}

% Rows where at least one method completed
\begin{table}[b]
\centering
\caption{\texttt{\detokenize{dj}}: cases where at least one method completed (incorrect results shown in bold).
For 16-, 32-, and 64-qubit shift4, the expected answer is ``Inequivalent.''
Both \textsc{ECMC} and \textsc{Manjushri} reported ``Equivalent,'' which is actually the correct answer because those circuits contain no occurrences of $R_z$, and thus no opportunity to inject a rotation error.
}
\label{tab:dj-partial}
\scriptsize
\setlength{\tabcolsep}{3.5pt}
\renewcommand{\arraystretch}{0.95}
\begin{tabular}{p{0.40\textwidth} r r r l r r}
\toprule
Circuit & $n$ & $|G|$ & $|G^\prime|$ & variant & \textsc{ECMC} & \textsc{Manjushri} \\
\midrule
\texttt{\detokenize{dj_nativegates_ibm_qiskit}} & 16 & 127 & 67 & flip & 0.67 & 0.07 \\
\texttt{\detokenize{dj_nativegates_ibm_qiskit}} & 16 & 127 & 66 & gm & 0.39 & 0.05 \\
\texttt{\detokenize{dj_nativegates_ibm_qiskit}} & 16 & 127 & 67 & opt & 2.01 & 0.13 \\
\texttt{\detokenize{dj_nativegates_ibm_qiskit}} & 16 & 127 & 67 & shift4 & \textbf{[0.80]} & \textbf{[0.13]} \\
\texttt{\detokenize{dj_nativegates_ibm_qiskit}} & 16 & 127 & 67 & shift7 & \textbf{0.78} & 0.09 \\
\texttt{\detokenize{dj_nativegates_ibm_qiskit}} & 32 & 249 & 129 & flip & 0.91 & \TO \\
\texttt{\detokenize{dj_nativegates_ibm_qiskit}} & 32 & 249 & 128 & gm & 0.72 & \TO \\
\texttt{\detokenize{dj_nativegates_ibm_qiskit}} & 32 & 249 & 129 & opt & 2.13 & 0.56 \\
\texttt{\detokenize{dj_nativegates_ibm_qiskit}} & 32 & 249 & 129 & shift4 & \textbf{[1.79]} & \textbf{[0.54]} \\
\texttt{\detokenize{dj_nativegates_ibm_qiskit}} & 32 & 249 & 129 & shift7 & \textbf{1.64} & \TO \\
\texttt{\detokenize{dj_nativegates_ibm_qiskit}} & 64 & 507 & 259 & flip & 1.40 & \TO \\
\texttt{\detokenize{dj_nativegates_ibm_qiskit}} & 64 & 507 & 258 & gm & 1.26 & \TO \\
\texttt{\detokenize{dj_nativegates_ibm_qiskit}} & 64 & 507 & 259 & opt & 4.68 & 3.14 \\
\texttt{\detokenize{dj_nativegates_ibm_qiskit}} & 64 & 507 & 259 & shift4 & \textbf{[3.94]} & \textbf{[2.97]} \\
\texttt{\detokenize{dj_nativegates_ibm_qiskit}} & 64 & 507 & 259 & shift7 & 6.91 & \TO \\
\bottomrule
\end{tabular}
\end{table}

% Rows where at least one method completed
\begin{table}[!tb]
\centering
\caption{\texttt{\detokenize{ghz}}: cases where at least one method completed (incorrect results shown in bold).}
\label{tab:ghz-partial}
\scriptsize
\setlength{\tabcolsep}{3.5pt}
\renewcommand{\arraystretch}{0.95}
\begin{tabular}{p{0.40\textwidth} r r r l r r}
\toprule
Circuit & $n$ & $|G|$ & $|G^\prime|$ & variant & \textsc{ECMC} & \textsc{Manjushri} \\
\midrule
\texttt{\detokenize{ghz_nativegates_ibm_qiskit}} & 16 & 18 & 45 & gm & 0.38 & 0.06 \\
\texttt{\detokenize{ghz_nativegates_ibm_qiskit}} & 16 & 18 & 46 & opt & 0.88 & 0.09 \\
\texttt{\detokenize{ghz_nativegates_ibm_qiskit}} & 32 & 34 & 93 & gm & 0.93 & \TO \\
\texttt{\detokenize{ghz_nativegates_ibm_qiskit}} & 32 & 34 & 94 & opt & 1.82 & 0.39 \\
\texttt{\detokenize{ghz_nativegates_ibm_qiskit}} & 64 & 66 & 189 & gm & 1.84 & \TO \\
\texttt{\detokenize{ghz_nativegates_ibm_qiskit}} & 64 & 66 & 190 & opt & 3.82 & 1.75 \\
\bottomrule
\end{tabular}
\end{table}

% Rows where at least one method completed
\begin{table}[!tb]
\centering
\caption{\texttt{\detokenize{graphstate}}: cases where at least one method completed (incorrect results shown in bold).}
\label{tab:graphstate-partial}
\scriptsize
\setlength{\tabcolsep}{3.5pt}
\renewcommand{\arraystretch}{0.95}
\begin{tabular}{p{0.40\textwidth} r r r l r r}
\toprule
Circuit & $n$ & $|G|$ & $|G^\prime|$ & variant & \textsc{ECMC} & \textsc{Manjushri} \\
\midrule
\texttt{\detokenize{graphstate_nativegates_ibm_qiskit}} & 16 & 160 & 31 & gm & 0.80 & 0.06 \\
\texttt{\detokenize{graphstate_nativegates_ibm_qiskit}} & 16 & 160 & 32 & opt & 1.04 & 0.15 \\
\texttt{\detokenize{graphstate_nativegates_ibm_qiskit}} & 32 & 320 & 63 & gm & 1.29 & 0.09 \\
\texttt{\detokenize{graphstate_nativegates_ibm_qiskit}} & 32 & 320 & 64 & opt & 1.97 & 0.18 \\
\texttt{\detokenize{graphstate_nativegates_ibm_qiskit}} & 64 & 640 & 127 & gm & 3.68 & 0.20 \\
\texttt{\detokenize{graphstate_nativegates_ibm_qiskit}} & 64 & 640 & 128 & opt & 8.34 & 0.43 \\
\bottomrule
\end{tabular}
\end{table}

% Rows where at least one method completed
\begin{table}[!tb]
\centering
\caption{\texttt{\detokenize{groundstate_medium}}: cases where at least one method completed (incorrect results shown in bold).}
\label{tab:groundstate-medium-partial}
\scriptsize
\setlength{\tabcolsep}{3.5pt}
\renewcommand{\arraystretch}{0.95}
\begin{tabular}{p{0.40\textwidth} r r r l r r}
\toprule
Circuit & $n$ & $|G|$ & $|G^\prime|$ & variant & \textsc{ECMC} & \textsc{Manjushri} \\
\midrule
\texttt{\detokenize{groundstate_medium_nativegates_qiskit}} & 12 & 1212 & 164 & flip & \TO & 1.55 \\
\texttt{\detokenize{groundstate_medium_nativegates_qiskit}} & 12 & 1212 & 163 & gm & \TO & 1.57 \\
\texttt{\detokenize{groundstate_medium_nativegates_qiskit}} & 12 & 1212 & 164 & shift4 & \TO & 1.28 \\
\texttt{\detokenize{groundstate_medium_nativegates_qiskit}} & 12 & 1212 & 164 & shift7 & \TO & 1.71 \\
\bottomrule
\end{tabular}
\end{table}

% Rows where at least one method completed
\begin{table}[!tb]
\centering
\caption{\texttt{\detokenize{groundstate_small}}: cases where at least one method completed (incorrect results shown in bold).}
\label{tab:groundstate-small-partial}
\scriptsize
\setlength{\tabcolsep}{3.5pt}
\renewcommand{\arraystretch}{0.95}
\begin{tabular}{p{0.40\textwidth} r r r l r r}
\toprule
Circuit & $n$ & $|G|$ & $|G^\prime|$ & variant & \textsc{ECMC} & \textsc{Manjushri} \\
\midrule
\texttt{\detokenize{groundstate_small_nativegates_qiskit}} & 4 & 180 & 35 & gm & 0.14 & 0.05 \\
\texttt{\detokenize{groundstate_small_nativegates_qiskit}} & 4 & 180 & 36 & opt & 0.39 & 0.09 \\
\texttt{\detokenize{groundstate_small_nativegates_qiskit}} & 4 & 180 & 36 & shift4 & 0.15 & 0.06 \\
\texttt{\detokenize{groundstate_small_nativegates_qiskit}} & 4 & 180 & 36 & shift7 & \textbf{0.43} & 0.04 \\
\bottomrule
\end{tabular}
\end{table}

% Rows where at least one method completed
\begin{table}[!tb]
\centering
\caption{\texttt{\detokenize{grover-noancilla}}: cases where at least one method completed (incorrect results shown in bold).}
\label{tab:grover-noancilla-partial}
\scriptsize
\setlength{\tabcolsep}{3.5pt}
\renewcommand{\arraystretch}{0.95}
\begin{tabular}{p{0.40\textwidth} r r r l r r}
\toprule
Circuit & $n$ & $|G|$ & $|G^\prime|$ & variant & \textsc{ECMC} & \textsc{Manjushri} \\
\midrule
\texttt{\detokenize{grover-noancilla_nativegates_ibm_qiskit}} & 3 & 190 & 190 & flip & 0.36 & 0.26 \\
\texttt{\detokenize{grover-noancilla_nativegates_ibm_qiskit}} & 3 & 190 & 189 & gm & 0.35 & 0.16 \\
\texttt{\detokenize{grover-noancilla_nativegates_ibm_qiskit}} & 3 & 190 & 190 & opt & 1.05 & 0.30 \\
\texttt{\detokenize{grover-noancilla_nativegates_ibm_qiskit}} & 3 & 190 & 190 & shift4 & 0.44 & 15.36 \\
\texttt{\detokenize{grover-noancilla_nativegates_ibm_qiskit}} & 3 & 190 & 190 & shift7 & \textbf{1.19} & 4.56 \\
\texttt{\detokenize{grover-noancilla_nativegates_ibm_qiskit}} & 4 & 499 & 629 & flip & 8.85 & 27.14 \\
\texttt{\detokenize{grover-noancilla_nativegates_ibm_qiskit}} & 4 & 499 & 628 & gm & 10.61 & 3.83 \\
\texttt{\detokenize{grover-noancilla_nativegates_ibm_qiskit}} & 4 & 499 & 629 & opt & 43.24 & \TO \\
\texttt{\detokenize{grover-noancilla_nativegates_ibm_qiskit}} & 4 & 499 & 629 & shift4 & 10.36 & \TO \\
\texttt{\detokenize{grover-noancilla_nativegates_ibm_qiskit}} & 4 & 499 & 629 & shift7 & \textbf{13.29} & \TO \\
\bottomrule
\end{tabular}
\end{table}

% Rows where at least one method completed
\begin{table}[!tb]
\centering
\caption{\texttt{\detokenize{grover-v-chain}}: cases where at least one method completed (incorrect results shown in bold).}
\label{tab:grover-v-chain-partial}
\scriptsize
\setlength{\tabcolsep}{3.5pt}
\renewcommand{\arraystretch}{0.95}
\begin{tabular}{p{0.40\textwidth} r r r l r r}
\toprule
Circuit & $n$ & $|G|$ & $|G^\prime|$ & variant & \textsc{ECMC} & \textsc{Manjushri} \\
\midrule
\texttt{\detokenize{grover-v-chain_nativegates_ibm_qiskit}} & 4 & 529 & 632 & flip & 29.98 & 24.58 \\
\texttt{\detokenize{grover-v-chain_nativegates_ibm_qiskit}} & 4 & 529 & 631 & gm & 30.07 & 7.51 \\
\texttt{\detokenize{grover-v-chain_nativegates_ibm_qiskit}} & 4 & 529 & 632 & shift4 & 26.58 & \TO \\
\bottomrule
\end{tabular}
\end{table}

\begin{table}[!tb]
\centering
\caption{\texttt{\detokenize{portfolioqaoa}}: cases where at least one method completed (incorrect results shown in bold).}
\label{tab:portfolioqaoa-partial}
\scriptsize
\setlength{\tabcolsep}{3.5pt}
\renewcommand{\arraystretch}{0.95}
\begin{tabular}{p{0.40\textwidth} r r r l r r}
\toprule
Circuit & $n$ & $|G|$ & $|G^\prime|$ & variant & \textsc{ECMC} & \textsc{Manjushri} \\
\midrule
\texttt{\detokenize{portfolioqaoa_nativegates_ibm_qiskit}} & 5 & 195 & 236 & flip & 7.33 & 32.65 \\
\texttt{\detokenize{portfolioqaoa_nativegates_ibm_qiskit}} & 5 & 195 & 235 & gm & 6.94 & 16.47 \\
\texttt{\detokenize{portfolioqaoa_nativegates_ibm_qiskit}} & 5 & 195 & 236 & opt & 9.60 & \TO \\
\texttt{\detokenize{portfolioqaoa_nativegates_ibm_qiskit}} & 5 & 195 & 236 & shift4 & 5.67 & 11.22 \\
\texttt{\detokenize{portfolioqaoa_nativegates_ibm_qiskit}} & 5 & 195 & 236 & shift7 & \textbf{12.30} & 18.72 \\
\texttt{\detokenize{portfolioqaoa_nativegates_ibm_qiskit}} & 6 & 261 & 356 & flip & 79.16 & \TO \\
\texttt{\detokenize{portfolioqaoa_nativegates_ibm_qiskit}} & 6 & 261 & 355 & gm & 73.19 & \TO \\
\texttt{\detokenize{portfolioqaoa_nativegates_ibm_qiskit}} & 6 & 261 & 356 & shift4 & 15.10 & \TO \\
\bottomrule
\end{tabular}
\end{table}

% Rows where at least one method completed
\begin{table}[!tb]
\centering
\caption{\texttt{\detokenize{portfoliovqe}}: cases where at least one method completed (incorrect results shown in bold).}
\label{tab:portfoliovqe-partial}
\scriptsize
\setlength{\tabcolsep}{3.5pt}
\renewcommand{\arraystretch}{0.95}
\begin{tabular}{p{0.40\textwidth} r r r l r r}
\toprule
Circuit & $n$ & $|G|$ & $|G^\prime|$ & variant & \textsc{ECMC} & \textsc{Manjushri} \\
\midrule
\texttt{\detokenize{portfoliovqe_nativegates_ibm_qiskit}} & 5 & 310 & 131 & flip & 6.64 & 4.68 \\
\texttt{\detokenize{portfoliovqe_nativegates_ibm_qiskit}} & 5 & 310 & 130 & gm & 5.51 & 5.63 \\
\texttt{\detokenize{portfoliovqe_nativegates_ibm_qiskit}} & 5 & 310 & 131 & opt & 8.27 & \TO \\
\texttt{\detokenize{portfoliovqe_nativegates_ibm_qiskit}} & 5 & 310 & 131 & shift4 & 6.58 & \TO \\
\texttt{\detokenize{portfoliovqe_nativegates_ibm_qiskit}} & 5 & 310 & 131 & shift7 & \textbf{8.17} & 84.18 \\
\texttt{\detokenize{portfoliovqe_nativegates_ibm_qiskit}} & 6 & 435 & 151 & flip & 41.77 & 34.62 \\
\texttt{\detokenize{portfoliovqe_nativegates_ibm_qiskit}} & 6 & 435 & 150 & gm & 35.18 & 52.99 \\
\texttt{\detokenize{portfoliovqe_nativegates_ibm_qiskit}} & 6 & 435 & 151 & shift4 & 46.68 & \TO \\
\bottomrule
\end{tabular}
\end{table}

% Rows where at least one method completed
\begin{table}[!tb]
\centering
\caption{\texttt{\detokenize{pricingcall}}: cases where at least one method completed (incorrect results shown in bold).}
\label{tab:pricingcall-partial}
\scriptsize
\setlength{\tabcolsep}{3.5pt}
\renewcommand{\arraystretch}{0.95}
\begin{tabular}{p{0.40\textwidth} r r r l r r}
\toprule
Circuit & $n$ & $|G|$ & $|G^\prime|$ & variant & \textsc{ECMC} & \textsc{Manjushri} \\
\midrule
\texttt{\detokenize{pricingcall_nativegates_ibm_qiskit}} & 5 & 240 & 166 & flip & 0.51 & 0.20 \\
\texttt{\detokenize{pricingcall_nativegates_ibm_qiskit}} & 5 & 240 & 165 & gm & 0.42 & 0.22 \\
\texttt{\detokenize{pricingcall_nativegates_ibm_qiskit}} & 5 & 240 & 166 & opt & 4.29 & 0.44 \\
\texttt{\detokenize{pricingcall_nativegates_ibm_qiskit}} & 5 & 240 & 166 & shift4 & 0.47 & 54.65 \\
\texttt{\detokenize{pricingcall_nativegates_ibm_qiskit}} & 5 & 240 & 166 & shift7 & \textbf{2.45} & \TO \\
\texttt{\detokenize{pricingcall_nativegates_ibm_qiskit}} & 7 & 422 & 277 & flip & 7.95 & 1.32 \\
\texttt{\detokenize{pricingcall_nativegates_ibm_qiskit}} & 7 & 422 & 276 & gm & 6.33 & 0.62 \\
\texttt{\detokenize{pricingcall_nativegates_ibm_qiskit}} & 7 & 422 & 277 & opt & 15.03 & 50.81 \\
\texttt{\detokenize{pricingcall_nativegates_ibm_qiskit}} & 7 & 422 & 277 & shift4 & 5.94 & \TO \\
\texttt{\detokenize{pricingcall_nativegates_ibm_qiskit}} & 7 & 422 & 277 & shift7 & \textbf{83.05} & \TO \\
\bottomrule
\end{tabular}
\end{table}

% Rows where at least one method completed
\begin{table}[!tb]
\centering
\caption{\texttt{\detokenize{pricingput}}: cases where at least one method completed (incorrect results shown in bold).}
\label{tab:pricingput-partial}
\scriptsize
\setlength{\tabcolsep}{3.5pt}
\renewcommand{\arraystretch}{0.95}
\begin{tabular}{p{0.40\textwidth} r r r l r r}
\toprule
Circuit & $n$ & $|G|$ & $|G^\prime|$ & variant & \textsc{ECMC} & \textsc{Manjushri} \\
\midrule
\texttt{\detokenize{pricingput_nativegates_ibm_qiskit}} & 5 & 240 & 192 & flip & 0.28 & 0.12 \\
\texttt{\detokenize{pricingput_nativegates_ibm_qiskit}} & 5 & 240 & 191 & gm & 0.24 & 0.13 \\
\texttt{\detokenize{pricingput_nativegates_ibm_qiskit}} & 5 & 240 & 192 & opt & 2.25 & 0.46 \\
\texttt{\detokenize{pricingput_nativegates_ibm_qiskit}} & 5 & 240 & 192 & shift4 & 1.34 & 0.43 \\
\texttt{\detokenize{pricingput_nativegates_ibm_qiskit}} & 5 & 240 & 192 & shift7 & \textbf{2.41} & 7.22 \\
\texttt{\detokenize{pricingput_nativegates_ibm_qiskit}} & 7 & 432 & 297 & flip & 19.65 & 1.74 \\
\texttt{\detokenize{pricingput_nativegates_ibm_qiskit}} & 7 & 432 & 296 & gm & 8.31 & 0.66 \\
\texttt{\detokenize{pricingput_nativegates_ibm_qiskit}} & 7 & 432 & 297 & opt & \TO & 10.35 \\
\texttt{\detokenize{pricingput_nativegates_ibm_qiskit}} & 7 & 432 & 297 & shift4 & 8.64 & \TO \\
\texttt{\detokenize{pricingput_nativegates_ibm_qiskit}} & 9 & 654 & 428 & opt & 19.78 & \TO \\
\bottomrule
\end{tabular}
\end{table}

% Rows where at least one method completed
\begin{table}[!tb]
\centering
\caption{\texttt{\detokenize{qaoa}}: cases where at least one method completed (incorrect results shown in bold).}
\label{tab:qaoa-partial}
\scriptsize
\setlength{\tabcolsep}{3.5pt}
\renewcommand{\arraystretch}{0.95}
\begin{tabular}{p{0.40\textwidth} r r r l r r}
\toprule
Circuit & $n$ & $|G|$ & $|G^\prime|$ & variant & \textsc{ECMC} & \textsc{Manjushri} \\
\midrule
\texttt{\detokenize{qaoa_nativegates_ibm_qiskit}} & 7 & 133 & 117 & flip & 0.33 & 0.17 \\
\texttt{\detokenize{qaoa_nativegates_ibm_qiskit}} & 7 & 133 & 116 & gm & 0.18 & 0.06 \\
\texttt{\detokenize{qaoa_nativegates_ibm_qiskit}} & 7 & 133 & 117 & opt & 0.58 & 0.23 \\
\texttt{\detokenize{qaoa_nativegates_ibm_qiskit}} & 7 & 133 & 117 & shift4 & 0.16 & 0.15 \\
\texttt{\detokenize{qaoa_nativegates_ibm_qiskit}} & 7 & 133 & 117 & shift7 & \textbf{0.55} & 0.15 \\
\texttt{\detokenize{qaoa_nativegates_ibm_qiskit}} & 9 & 171 & 296 & flip & 0.35 & 0.22 \\
\texttt{\detokenize{qaoa_nativegates_ibm_qiskit}} & 9 & 171 & 295 & gm & 0.36 & \TO \\
\texttt{\detokenize{qaoa_nativegates_ibm_qiskit}} & 9 & 171 & 296 & opt & 0.97 & 2.30 \\
\texttt{\detokenize{qaoa_nativegates_ibm_qiskit}} & 9 & 171 & 296 & shift4 & 0.40 & 2.73 \\
\texttt{\detokenize{qaoa_nativegates_ibm_qiskit}} & 9 & 171 & 296 & shift7 & \textbf{1.08} & \TO \\
\texttt{\detokenize{qaoa_nativegates_ibm_qiskit}} & 11 & 209 & 359 & flip & 0.71 & 1.54 \\
\texttt{\detokenize{qaoa_nativegates_ibm_qiskit}} & 11 & 209 & 358 & gm & 0.36 & 0.55 \\
\texttt{\detokenize{qaoa_nativegates_ibm_qiskit}} & 11 & 209 & 359 & opt & 1.01 & 2.28 \\
\texttt{\detokenize{qaoa_nativegates_ibm_qiskit}} & 11 & 209 & 359 & shift4 & 0.31 & 0.65 \\
\texttt{\detokenize{qaoa_nativegates_ibm_qiskit}} & 11 & 209 & 359 & shift7 & \textbf{1.27} & 0.33 \\
\texttt{\detokenize{qaoa_nativegates_ibm_qiskit}} & 14 & 266 & 280 & flip & 0.44 & 0.08 \\
\texttt{\detokenize{qaoa_nativegates_ibm_qiskit}} & 14 & 266 & 279 & gm & 0.42 & 0.53 \\
\texttt{\detokenize{qaoa_nativegates_ibm_qiskit}} & 14 & 266 & 280 & opt & 1.20 & 0.97 \\
\texttt{\detokenize{qaoa_nativegates_ibm_qiskit}} & 14 & 266 & 280 & shift4 & 0.42 & 0.50 \\
\texttt{\detokenize{qaoa_nativegates_ibm_qiskit}} & 14 & 266 & 280 & shift7 & 1.29 & 0.33 \\
\texttt{\detokenize{qaoa_nativegates_ibm_qiskit}} & 15 & 285 & 587 & flip & 0.48 & 0.87 \\
\texttt{\detokenize{qaoa_nativegates_ibm_qiskit}} & 15 & 285 & 586 & gm & 0.59 & 3.11 \\
\texttt{\detokenize{qaoa_nativegates_ibm_qiskit}} & 15 & 285 & 587 & opt & 2.59 & 5.61 \\
\texttt{\detokenize{qaoa_nativegates_ibm_qiskit}} & 15 & 285 & 587 & shift4 & 0.62 & 0.44 \\
\texttt{\detokenize{qaoa_nativegates_ibm_qiskit}} & 15 & 285 & 587 & shift7 & \textbf{2.71} & 0.56 \\
\texttt{\detokenize{qaoa_nativegates_ibm_qiskit}} & 16 & 304 & 461 & flip & 0.56 & \TO \\
\texttt{\detokenize{qaoa_nativegates_ibm_qiskit}} & 16 & 304 & 460 & gm & 0.45 & 0.98 \\
\texttt{\detokenize{qaoa_nativegates_ibm_qiskit}} & 16 & 304 & 461 & opt & 1.42 & 4.38 \\
\texttt{\detokenize{qaoa_nativegates_ibm_qiskit}} & 16 & 304 & 461 & shift4 & 0.91 & 69.36 \\
\texttt{\detokenize{qaoa_nativegates_ibm_qiskit}} & 16 & 304 & 461 & shift7 & \textbf{1.58} & 2.00 \\
\bottomrule
\end{tabular}
\end{table}

% Rows where at least one method completed
\begin{table}[!tb]
\centering
\caption{\texttt{\detokenize{qft}}: cases where at least one method completed (incorrect results shown in bold).}
\label{tab:qft-partial}
\scriptsize
\setlength{\tabcolsep}{3.5pt}
\renewcommand{\arraystretch}{0.95}
\begin{tabular}{p{0.40\textwidth} r r r l r r}
\toprule
Circuit & $n$ & $|G|$ & $|G^\prime|$ & variant & \textsc{ECMC} & \textsc{Manjushri} \\
\midrule
\texttt{\detokenize{qft_nativegates_ibm_qiskit}} & 2 & 14 & 14 & flip & 0.11 & 0.03 \\
\texttt{\detokenize{qft_nativegates_ibm_qiskit}} & 2 & 14 & 13 & gm & 0.10 & 0.02 \\
\texttt{\detokenize{qft_nativegates_ibm_qiskit}} & 2 & 14 & 14 & opt & 0.10 & 0.03 \\
\texttt{\detokenize{qft_nativegates_ibm_qiskit}} & 2 & 14 & 14 & shift4 & 0.10 & 0.03 \\
\texttt{\detokenize{qft_nativegates_ibm_qiskit}} & 2 & 14 & 14 & shift7 & \textbf{0.10} & 0.03 \\
\texttt{\detokenize{qft_nativegates_ibm_qiskit}} & 8 & 176 & 228 & flip & 0.24 & 0.27 \\
\texttt{\detokenize{qft_nativegates_ibm_qiskit}} & 8 & 176 & 227 & gm & 0.24 & 0.17 \\
\texttt{\detokenize{qft_nativegates_ibm_qiskit}} & 8 & 176 & 228 & opt & 3.40 & 1.88 \\
\texttt{\detokenize{qft_nativegates_ibm_qiskit}} & 8 & 176 & 228 & shift4 & 0.30 & \TO \\
\texttt{\detokenize{qft_nativegates_ibm_qiskit}} & 8 & 176 & 228 & shift7 & 1.38 & \TO \\
\texttt{\detokenize{qft_nativegates_ibm_qiskit}} & 16 & 672 & 814 & flip & 1.50 & \TO \\
\texttt{\detokenize{qft_nativegates_ibm_qiskit}} & 16 & 672 & 813 & gm & 0.83 & \TO \\
\texttt{\detokenize{qft_nativegates_ibm_qiskit}} & 16 & 672 & 814 & opt & 1.73 & \TO \\
\texttt{\detokenize{qft_nativegates_ibm_qiskit}} & 16 & 672 & 814 & shift4 & 1.15 & \TO \\
\texttt{\detokenize{qft_nativegates_ibm_qiskit}} & 32 & 2624 & 3176 & gm & \textbf{73.01} & \TO \\
\texttt{\detokenize{qft_nativegates_ibm_qiskit}} & 32 & 2624 & 3177 & opt & 70.83 & \TO \\
\texttt{\detokenize{qft_nativegates_ibm_qiskit}} & 32 & 2624 & 3177 & shift4 & \textbf{72.25} & \TO \\
\texttt{\detokenize{qft_nativegates_ibm_qiskit}} & 32 & 2624 & 3177 & shift7 & 71.32 & \TO \\
\bottomrule
\end{tabular}
\end{table}

% Rows where at least one method completed
\begin{table}[!tb]
\centering
\caption{\texttt{\detokenize{qftentangled}}: cases where at least one method completed (incorrect results shown in bold).}
\label{tab:qftentangled-partial}
\scriptsize
\setlength{\tabcolsep}{3.5pt}
\renewcommand{\arraystretch}{0.95}
\begin{tabular}{p{0.40\textwidth} r r r l r r}
\toprule
Circuit & $n$ & $|G|$ & $|G^\prime|$ & variant & \textsc{ECMC} & \textsc{Manjushri} \\
\midrule
\texttt{\detokenize{qftentangled_nativegates_ibm_qiskit}} & 16 & 690 & 853 & flip & 0.61 & \TO \\
\texttt{\detokenize{qftentangled_nativegates_ibm_qiskit}} & 16 & 690 & 852 & gm & 51.73 & \TO \\
\texttt{\detokenize{qftentangled_nativegates_ibm_qiskit}} & 16 & 690 & 853 & opt & 51.60 & \TO \\
\texttt{\detokenize{qftentangled_nativegates_ibm_qiskit}} & 16 & 690 & 853 & shift4 & 0.82 & \TO \\
\bottomrule
\end{tabular}
\end{table}

% Rows where at least one method completed
\begin{table}[!tb]
\centering
\caption{\texttt{\detokenize{qnn}}: cases where at least one method completed (incorrect results shown in bold).}
\label{tab:qnn-partial}
\scriptsize
\setlength{\tabcolsep}{3.5pt}
\renewcommand{\arraystretch}{0.95}
\begin{tabular}{p{0.40\textwidth} r r r l r r}
\toprule
Circuit & $n$ & $|G|$ & $|G^\prime|$ & variant & \textsc{ECMC} & \textsc{Manjushri} \\
\midrule
\texttt{\detokenize{qnn_nativegates_ibm_qiskit}} & 2 & 43 & 36 & flip & 0.11 & 0.03 \\
\texttt{\detokenize{qnn_nativegates_ibm_qiskit}} & 2 & 43 & 35 & gm & 0.10 & 0.03 \\
\texttt{\detokenize{qnn_nativegates_ibm_qiskit}} & 2 & 43 & 36 & opt & 0.13 & 0.03 \\
\texttt{\detokenize{qnn_nativegates_ibm_qiskit}} & 2 & 43 & 36 & shift4 & 0.10 & 0.04 \\
\texttt{\detokenize{qnn_nativegates_ibm_qiskit}} & 2 & 43 & 36 & shift7 & \textbf{0.14} & 0.03 \\
\texttt{\detokenize{qnn_nativegates_ibm_qiskit}} & 8 & 319 & 494 & flip & 0.40 & \TO \\
\texttt{\detokenize{qnn_nativegates_ibm_qiskit}} & 32 & 3583 & 8216 & shift4 & 8.39 & \TO \\
\texttt{\detokenize{qnn_nativegates_ibm_qiskit}} & 32 & 3583 & 8216 & shift7 & 7.31 & \TO \\
\bottomrule
\end{tabular}
\end{table}

% Rows where at least one method completed
\begin{table}[!tb]
\centering
\caption{\texttt{\detokenize{qpeexact}}: cases where at least one method completed (incorrect results shown in bold).}
\label{tab:qpeexact-partial}
\scriptsize
\setlength{\tabcolsep}{3.5pt}
\renewcommand{\arraystretch}{0.95}
\begin{tabular}{p{0.40\textwidth} r r r l r r}
\toprule
Circuit & $n$ & $|G|$ & $|G^\prime|$ & variant & \textsc{ECMC} & \textsc{Manjushri} \\
\midrule
\texttt{\detokenize{qpeexact_nativegates_ibm_qiskit}} & 15 & 712 & 837 & flip & 1.05 & \TO \\
\texttt{\detokenize{qpeexact_nativegates_ibm_qiskit}} & 15 & 712 & 836 & gm & 1.31 & \TO \\
\texttt{\detokenize{qpeexact_nativegates_ibm_qiskit}} & 15 & 712 & 837 & opt & 68.48 & \TO \\
\texttt{\detokenize{qpeexact_nativegates_ibm_qiskit}} & 15 & 712 & 837 & shift4 & 0.88 & \TO \\
\texttt{\detokenize{qpeexact_nativegates_ibm_qiskit}} & 31 & 2712 & 3276 & flip & 7.18 & \TO \\
\texttt{\detokenize{qpeexact_nativegates_ibm_qiskit}} & 31 & 2712 & 3276 & shift4 & \textbf{7.94} & \TO \\
\bottomrule
\end{tabular}
\end{table}

% Rows where at least one method completed
\begin{table}[!tb]
\centering
\caption{\texttt{\detokenize{qpeinexact}}: cases where at least one method completed (incorrect results shown in bold).}
\label{tab:qpeinexact-partial}
\scriptsize
\setlength{\tabcolsep}{3.5pt}
\renewcommand{\arraystretch}{0.95}
\begin{tabular}{p{0.40\textwidth} r r r l r r}
\toprule
Circuit & $n$ & $|G|$ & $|G^\prime|$ & variant & \textsc{ECMC} & \textsc{Manjushri} \\
\midrule
\texttt{\detokenize{qpeinexact_nativegates_ibm_qiskit}} & 15 & 712 & 848 & flip & 3.21 & \TO \\
\texttt{\detokenize{qpeinexact_nativegates_ibm_qiskit}} & 15 & 712 & 847 & gm & 4.50 & \TO \\
\texttt{\detokenize{qpeinexact_nativegates_ibm_qiskit}} & 15 & 712 & 848 & opt & 11.07 & \TO \\
\texttt{\detokenize{qpeinexact_nativegates_ibm_qiskit}} & 15 & 712 & 848 & shift4 & 0.72 & \TO \\
\texttt{\detokenize{qpeinexact_nativegates_ibm_qiskit}} & 31 & 2712 & 3179 & flip & 6.27 & \TO \\
\bottomrule
\end{tabular}
\end{table}

% Rows where at least one method completed
\begin{table}[!tb]
\centering
\caption{\texttt{\detokenize{rd84_313}}: cases where at least one method completed (incorrect results shown in bold).}
\label{tab:rd84-313-partial}
\scriptsize
\setlength{\tabcolsep}{3.5pt}
\renewcommand{\arraystretch}{0.95}
\begin{tabular}{p{0.40\textwidth} r r r l r r}
\toprule
Circuit & $n$ & $|G|$ & $|G^\prime|$ & variant & \textsc{ECMC} & \textsc{Manjushri} \\
\midrule
\texttt{\detokenize{rd84_313}} & 34 & 804 & 1233 & flip & 1.32 & \TO \\
\texttt{\detokenize{rd84_313}} & 34 & 804 & 1292 & gm & \TO & 55.93 \\
\bottomrule
\end{tabular}
\end{table}

% Rows where at least one method completed
\begin{table}[!tb]
\centering
\caption{\texttt{\detokenize{realamprandom}}: cases where at least one method completed (incorrect results shown in bold).}
\label{tab:realamprandom-partial}
\scriptsize
\setlength{\tabcolsep}{3.5pt}
\renewcommand{\arraystretch}{0.95}
\begin{tabular}{p{0.40\textwidth} r r r l r r}
\toprule
Circuit & $n$ & $|G|$ & $|G^\prime|$ & variant & \textsc{ECMC} & \textsc{Manjushri} \\
\midrule
\texttt{\detokenize{realamprandom_nativegates_ibm_qiskit}} & 16 & 680 & 678 & gm & 1.15 & \TO \\
\texttt{\detokenize{realamprandom_nativegates_ibm_qiskit}} & 16 & 680 & 679 & opt & 2.26 & \TO \\
\texttt{\detokenize{realamprandom_nativegates_ibm_qiskit}} & 32 & 2128 & 2215 & flip & 4.93 & \TO \\
\texttt{\detokenize{realamprandom_nativegates_ibm_qiskit}} & 32 & 2128 & 2215 & opt & 25.43 & \TO \\
\texttt{\detokenize{realamprandom_nativegates_ibm_qiskit}} & 32 & 2128 & 2215 & shift4 & \textbf{25.89} & \TO \\
\texttt{\detokenize{realamprandom_nativegates_ibm_qiskit}} & 64 & 7328 & 7410 & gm & 21.54 & \TO \\
\texttt{\detokenize{realamprandom_nativegates_ibm_qiskit}} & 64 & 7328 & 7411 & opt & 28.10 & \TO \\
\bottomrule
\end{tabular}
\end{table}

% Rows where at least one method completed
\begin{table}[!tb]
\centering
\caption{\texttt{\detokenize{routing}}: cases where at least one method completed (incorrect results shown in bold).}
\label{tab:routing-partial}
\scriptsize
\setlength{\tabcolsep}{3.5pt}
\renewcommand{\arraystretch}{0.95}
\begin{tabular}{p{0.40\textwidth} r r r l r r}
\toprule
Circuit & $n$ & $|G|$ & $|G^\prime|$ & variant & \textsc{ECMC} & \textsc{Manjushri} \\
\midrule
\texttt{\detokenize{routing_nativegates_ibm_qiskit}} & 2 & 43 & 29 & flip & 0.10 & 0.03 \\
\texttt{\detokenize{routing_nativegates_ibm_qiskit}} & 2 & 43 & 28 & gm & 0.11 & 0.03 \\
\texttt{\detokenize{routing_nativegates_ibm_qiskit}} & 2 & 43 & 29 & opt & 0.14 & 0.04 \\
\texttt{\detokenize{routing_nativegates_ibm_qiskit}} & 2 & 43 & 29 & shift4 & 0.10 & 0.03 \\
\texttt{\detokenize{routing_nativegates_ibm_qiskit}} & 2 & 43 & 29 & shift7 & \textbf{0.13} & 0.03 \\
\texttt{\detokenize{routing_nativegates_ibm_qiskit}} & 6 & 135 & 142 & flip & 0.40 & 0.23 \\
\texttt{\detokenize{routing_nativegates_ibm_qiskit}} & 6 & 135 & 141 & gm & 0.39 & 70.65 \\
\texttt{\detokenize{routing_nativegates_ibm_qiskit}} & 6 & 135 & 142 & opt & 20.91 & \TO \\
\texttt{\detokenize{routing_nativegates_ibm_qiskit}} & 6 & 135 & 142 & shift4 & 0.34 & \TO \\
\texttt{\detokenize{routing_nativegates_ibm_qiskit}} & 6 & 135 & 142 & shift7 & \textbf{5.77} & \TO \\
\texttt{\detokenize{routing_nativegates_ibm_qiskit}} & 12 & 273 & 409 & flip & 0.45 & \TO \\
\texttt{\detokenize{routing_nativegates_ibm_qiskit}} & 12 & 273 & 409 & opt & 4.43 & \TO \\
\texttt{\detokenize{routing_nativegates_ibm_qiskit}} & 12 & 273 & 409 & shift4 & 6.48 & \TO \\
\texttt{\detokenize{routing_nativegates_ibm_qiskit}} & 12 & 273 & 409 & shift7 & \textbf{18.02} & \TO \\
\bottomrule
\end{tabular}
\end{table}

% Rows where at least one method completed
\begin{table}[!tb]
\centering
\caption{\texttt{\detokenize{su2random}}: cases where at least one method completed (incorrect results shown in bold).}
\label{tab:su2random-partial}
\scriptsize
\setlength{\tabcolsep}{3.5pt}
\renewcommand{\arraystretch}{0.95}
\begin{tabular}{p{0.40\textwidth} r r r l r r}
\toprule
Circuit & $n$ & $|G|$ & $|G^\prime|$ & variant & \textsc{ECMC} & \textsc{Manjushri} \\
\midrule
\texttt{\detokenize{su2random_nativegates_ibm_qiskit}} & 4 & 114 & 89 & gm & 0.14 & 0.07 \\
\texttt{\detokenize{su2random_nativegates_ibm_qiskit}} & 4 & 114 & 90 & opt & 0.86 & 0.48 \\
\texttt{\detokenize{su2random_nativegates_ibm_qiskit}} & 4 & 114 & 90 & shift4 & 0.13 & 0.32 \\
\texttt{\detokenize{su2random_nativegates_ibm_qiskit}} & 4 & 114 & 90 & shift7 & \textbf{0.63} & 0.16 \\
\texttt{\detokenize{su2random_nativegates_ibm_qiskit}} & 8 & 276 & 185 & gm & 1.67 & 6.14 \\
\texttt{\detokenize{su2random_nativegates_ibm_qiskit}} & 8 & 276 & 186 & shift4 & 41.97 & \TO \\
\texttt{\detokenize{su2random_nativegates_ibm_qiskit}} & 16 & 744 & 377 & gm & 3.34 & \TO \\
\texttt{\detokenize{su2random_nativegates_ibm_qiskit}} & 16 & 744 & 378 & opt & 10.27 & \TO \\
\texttt{\detokenize{su2random_nativegates_ibm_qiskit}} & 17 & 816 & 401 & gm & 2.60 & \TO \\
\texttt{\detokenize{su2random_nativegates_ibm_qiskit}} & 17 & 816 & 402 & opt & 2.69 & \TO \\
\texttt{\detokenize{su2random_nativegates_ibm_qiskit}} & 17 & 816 & 402 & shift4 & 10.52 & \TO \\
\texttt{\detokenize{su2random_nativegates_ibm_qiskit}} & 19 & 969 & 450 & shift4 & 1.09 & \TO \\
\texttt{\detokenize{su2random_nativegates_ibm_qiskit}} & 19 & 969 & 450 & shift7 & \textbf{35.51} & \TO \\
\texttt{\detokenize{su2random_nativegates_ibm_qiskit}} & 32 & 2256 & 762 & shift4 & \textbf{21.41} & \TO \\
\texttt{\detokenize{su2random_nativegates_ibm_qiskit}} & 32 & 2256 & 762 & shift7 & 4.87 & \TO \\
\bottomrule
\end{tabular}
\end{table}

% Rows where at least one method completed
\begin{table}[!tb]
\centering
\caption{\texttt{\detokenize{tsp}}: cases where at least one method completed (incorrect results shown in bold).}
\label{tab:tsp-partial}
\scriptsize
\setlength{\tabcolsep}{3.5pt}
\renewcommand{\arraystretch}{0.95}
\begin{tabular}{p{0.40\textwidth} r r r l r r}
\toprule
Circuit & $n$ & $|G|$ & $|G^\prime|$ & variant & \textsc{ECMC} & \textsc{Manjushri} \\
\midrule
\texttt{\detokenize{tsp_nativegates_ibm_qiskit}} & 4 & 225 & 86 & flip & 0.64 & 0.08 \\
\texttt{\detokenize{tsp_nativegates_ibm_qiskit}} & 4 & 225 & 85 & gm & 0.71 & 0.09 \\
\texttt{\detokenize{tsp_nativegates_ibm_qiskit}} & 4 & 225 & 86 & opt & 2.14 & 0.52 \\
\texttt{\detokenize{tsp_nativegates_ibm_qiskit}} & 4 & 225 & 86 & shift4 & 0.57 & 0.13 \\
\texttt{\detokenize{tsp_nativegates_ibm_qiskit}} & 4 & 225 & 86 & shift7 & \textbf{2.05} & 0.13 \\
\texttt{\detokenize{tsp_nativegates_ibm_qiskit}} & 9 & 550 & 315 & flip & \TO & 0.40 \\
\texttt{\detokenize{tsp_nativegates_ibm_qiskit}} & 9 & 550 & 314 & gm & \TO & 0.64 \\
\texttt{\detokenize{tsp_nativegates_ibm_qiskit}} & 16 & 1005 & 623 & flip & 0.64 & \TO \\
\texttt{\detokenize{tsp_nativegates_ibm_qiskit}} & 16 & 1005 & 622 & gm & 37.15 & \TO \\
\bottomrule
\end{tabular}
\end{table}

% Rows where at least one method completed
\begin{table}[!tb]
\centering
\caption{\texttt{\detokenize{twolocalrandom}}: cases where at least one method completed (incorrect results shown in bold).}
\label{tab:twolocalrandom-partial}
\scriptsize
\setlength{\tabcolsep}{3.5pt}
\renewcommand{\arraystretch}{0.95}
\begin{tabular}{p{0.40\textwidth} r r r l r r}
\toprule
Circuit & $n$ & $|G|$ & $|G^\prime|$ & variant & \textsc{ECMC} & \textsc{Manjushri} \\
\midrule
\texttt{\detokenize{twolocalrandom_nativegates_ibm_qiskit}} & 16 & 680 & 679 & flip & \textbf{56.93} & \TO \\
\texttt{\detokenize{twolocalrandom_nativegates_ibm_qiskit}} & 16 & 680 & 678 & gm & 0.99 & \TO \\
\texttt{\detokenize{twolocalrandom_nativegates_ibm_qiskit}} & 32 & 2128 & 2215 & flip & 4.97 & \TO \\
\texttt{\detokenize{twolocalrandom_nativegates_ibm_qiskit}} & 32 & 2128 & 2214 & gm & 5.00 & \TO \\
\texttt{\detokenize{twolocalrandom_nativegates_ibm_qiskit}} & 32 & 2128 & 2215 & opt & 24.31 & \TO \\
\texttt{\detokenize{twolocalrandom_nativegates_ibm_qiskit}} & 64 & 7328 & 7411 & opt & 32.27 & \TO \\
\bottomrule
\end{tabular}
\end{table}

% Rows where at least one method completed
\begin{table}[!tb]
\centering
\caption{\texttt{\detokenize{vqe}}: cases where at least one method completed (incorrect results shown in bold).}
\label{tab:vqe-partial}
\scriptsize
\setlength{\tabcolsep}{3.5pt}
\renewcommand{\arraystretch}{0.95}
\begin{tabular}{p{0.40\textwidth} r r r l r r}
\toprule
Circuit & $n$ & $|G|$ & $|G^\prime|$ & variant & \textsc{ECMC} & \textsc{Manjushri} \\
\midrule
\texttt{\detokenize{vqe_nativegates_ibm_qiskit}} & 5 & 83 & 83 & flip & 0.14 & 0.03 \\
\texttt{\detokenize{vqe_nativegates_ibm_qiskit}} & 5 & 83 & 82 & gm & 0.13 & 0.04 \\
\texttt{\detokenize{vqe_nativegates_ibm_qiskit}} & 5 & 83 & 83 & opt & 0.48 & 83.52 \\
\texttt{\detokenize{vqe_nativegates_ibm_qiskit}} & 5 & 83 & 83 & shift4 & 0.15 & 0.87 \\
\texttt{\detokenize{vqe_nativegates_ibm_qiskit}} & 5 & 83 & 83 & shift7 & \textbf{0.36} & 0.40 \\
\texttt{\detokenize{vqe_nativegates_ibm_qiskit}} & 10 & 168 & 221 & flip & 0.28 & 0.06 \\
\texttt{\detokenize{vqe_nativegates_ibm_qiskit}} & 10 & 168 & 220 & gm & 0.79 & \TO \\
\texttt{\detokenize{vqe_nativegates_ibm_qiskit}} & 10 & 168 & 221 & opt & 1.70 & \TO \\
\texttt{\detokenize{vqe_nativegates_ibm_qiskit}} & 10 & 168 & 221 & shift4 & 0.87 & \TO \\
\texttt{\detokenize{vqe_nativegates_ibm_qiskit}} & 10 & 168 & 221 & shift7 & \textbf{1.52} & \TO \\
\texttt{\detokenize{vqe_nativegates_ibm_qiskit}} & 14 & 236 & 327 & flip & 0.45 & 0.23 \\
\texttt{\detokenize{vqe_nativegates_ibm_qiskit}} & 14 & 236 & 326 & gm & 0.44 & 0.89 \\
\texttt{\detokenize{vqe_nativegates_ibm_qiskit}} & 14 & 236 & 327 & opt & 1.93 & \TO \\
\texttt{\detokenize{vqe_nativegates_ibm_qiskit}} & 14 & 236 & 327 & shift4 & 0.47 & \TO \\
\texttt{\detokenize{vqe_nativegates_ibm_qiskit}} & 14 & 236 & 327 & shift7 & \textbf{4.56} & 1.61 \\
\texttt{\detokenize{vqe_nativegates_ibm_qiskit}} & 15 & 253 & 349 & flip & 0.47 & 0.16 \\
\texttt{\detokenize{vqe_nativegates_ibm_qiskit}} & 15 & 253 & 348 & gm & 0.46 & 0.54 \\
\texttt{\detokenize{vqe_nativegates_ibm_qiskit}} & 15 & 253 & 349 & opt & 9.19 & \TO \\
\texttt{\detokenize{vqe_nativegates_ibm_qiskit}} & 15 & 253 & 349 & shift4 & 0.63 & \TO \\
\texttt{\detokenize{vqe_nativegates_ibm_qiskit}} & 15 & 253 & 349 & shift7 & \textbf{5.23} & \TO \\
\texttt{\detokenize{vqe_nativegates_ibm_qiskit}} & 16 & 270 & 424 & flip & 1.69 & \TO \\
\texttt{\detokenize{vqe_nativegates_ibm_qiskit}} & 16 & 270 & 423 & gm & 0.51 & 4.28 \\
\texttt{\detokenize{vqe_nativegates_ibm_qiskit}} & 16 & 270 & 424 & opt & 7.03 & \TO \\
\texttt{\detokenize{vqe_nativegates_ibm_qiskit}} & 16 & 270 & 424 & shift4 & 2.45 & \TO \\
\texttt{\detokenize{vqe_nativegates_ibm_qiskit}} & 16 & 270 & 424 & shift7 & \textbf{8.75} & \TO \\
\bottomrule
\end{tabular}
\end{table}

% Rows where at least one method completed
\begin{table}[!tb]
\centering
\caption{\texttt{\detokenize{wstate}}: cases where at least one method completed (incorrect results shown in bold).}
\label{tab:wstate-partial}
\scriptsize
\setlength{\tabcolsep}{3.5pt}
\renewcommand{\arraystretch}{0.95}
\begin{tabular}{p{0.40\textwidth} r r r l r r}
\toprule
Circuit & $n$ & $|G|$ & $|G^\prime|$ & variant & \textsc{ECMC} & \textsc{Manjushri} \\
\midrule
\texttt{\detokenize{wstate_nativegates_ibm_qiskit}} & 16 & 271 & 242 & flip & 0.73 & 0.17 \\
\texttt{\detokenize{wstate_nativegates_ibm_qiskit}} & 16 & 271 & 241 & gm & 0.98 & 0.29 \\
\texttt{\detokenize{wstate_nativegates_ibm_qiskit}} & 16 & 271 & 242 & opt & 1.65 & 0.44 \\
\texttt{\detokenize{wstate_nativegates_ibm_qiskit}} & 16 & 271 & 242 & shift4 & 0.59 & 0.15 \\
\texttt{\detokenize{wstate_nativegates_ibm_qiskit}} & 16 & 271 & 242 & shift7 & \textbf{1.48} & 0.10 \\
\texttt{\detokenize{wstate_nativegates_ibm_qiskit}} & 32 & 559 & 498 & flip & 1.32 & \TO \\
\texttt{\detokenize{wstate_nativegates_ibm_qiskit}} & 32 & 559 & 497 & gm & 2.72 & \TO \\
\texttt{\detokenize{wstate_nativegates_ibm_qiskit}} & 32 & 559 & 498 & opt & 4.90 & 3.70 \\
\texttt{\detokenize{wstate_nativegates_ibm_qiskit}} & 32 & 559 & 498 & shift4 & 1.97 & \TO \\
\texttt{\detokenize{wstate_nativegates_ibm_qiskit}} & 32 & 559 & 498 & shift7 & \textbf{5.24} & \TO \\
\texttt{\detokenize{wstate_nativegates_ibm_qiskit}} & 64 & 1135 & 1010 & flip & 4.77 & \TO \\
\texttt{\detokenize{wstate_nativegates_ibm_qiskit}} & 64 & 1135 & 1009 & gm & 9.21 & \TO \\
\texttt{\detokenize{wstate_nativegates_ibm_qiskit}} & 64 & 1135 & 1010 & opt & 19.49 & 39.38 \\
\texttt{\detokenize{wstate_nativegates_ibm_qiskit}} & 64 & 1135 & 1010 & shift4 & 4.46 & \TO \\
\texttt{\detokenize{wstate_nativegates_ibm_qiskit}} & 64 & 1135 & 1010 & shift7 & 4.79 & \TO \\
\bottomrule
\end{tabular}
\end{table}

\end{document}